\documentclass[aps, prd, onecolumn, tightenlines, notitlepage, superscriptaddress, nofootinbib, preprintnumbers, floatfix,showkeys,11pt,altaffilletter]{revtex4-1}
\usepackage{upgreek}
\usepackage[official]{eurosym}
\usepackage[normalem]{ulem}
\usepackage{amstext}
\usepackage{graphicx}
\graphicspath{{figures/}}
\usepackage{url}
\usepackage{color}
\usepackage{ulem}
\usepackage[version=4]{mhchem}
\usepackage[utf8]{inputenc}
\usepackage{fontawesome}
\usepackage{yfonts}
\usepackage{mathrsfs}
\pdfoutput=1
\usepackage{textcomp}
\usepackage{comment}
\usepackage{yfonts}
\usepackage{epsfig,amsfonts,mathrsfs,graphicx,color,slashed,multirow}
\usepackage{latexsym,graphicx,slashed,color,enumerate,url,cancel,gensymb}
\usepackage{textcomp}

\usepackage[x11names]{xcolor}
\usepackage[colorlinks,pdfstartview=FitV,breaklinks=true]{hyperref}
\usepackage{booktabs}
\usepackage{adjustbox}
\usepackage{textgreek} 
\usepackage{caption, subcaption} 
\usepackage{booktabs} 
\usepackage{float}
\usepackage{lmodern}
\usepackage{ae,aecompl}
\usepackage{appendix}
\usepackage{orcidlink}
\usepackage{nccmath} 
\usepackage{textcomp}
\allowdisplaybreaks 


\def\cevns{CE\textnu NS}
\def\eves{E\textnu ES}
\def\d{\mathrm{d}}
\newcommand{\qtransfer}{\left|\mathbf{q}\right|}

\definecolor{amber}{rgb}{1.0, 0.49, 0.0}

\definecolor{magenta(dye)}{rgb}{0.79, 0.08, 0.48}
\definecolor{byzantium}{rgb}{0.44, 0.16, 0.39}

\DeclareMathOperator{\diag}{diag}

\makeatletter
    \newcommand{\colorboxed}[3][white]{\fcolorbox{#2}{#1}{\m@th$\displaystyle#3$}}
\makeatother

\AtBeginDocument{\hypersetup{citecolor=magenta(dye),linkcolor=magenta(dye),urlcolor=magenta(dye)}}
\usepackage{appendix}

\begin{document}

\sloppy

\title{{\Large Invisible decay of solar neutrinos at dark matter experiments}}

\author{Martina Beccaria~\orcidlink{0009-0007-1124-8262}}
\email{martina.beccaria@gssi.it}
\affiliation{Gran Sasso Science Institute, Viale F. Crispi 7, L’Aquila, 67100, Italy}
\affiliation{Istituto Nazionale di Fisica Nucleare (INFN), Laboratori Nazionali del Gran Sasso, 67100 Assergi, L’Aquila (AQ), Italy
}

\author{Veronica Beligotti~\orcidlink{0009-0000-9097-472X}}
\email{veronica.beligotti@gssi.it}
\affiliation{Gran Sasso Science Institute, Viale F. Crispi 7, L’Aquila, 67100, Italy}
\affiliation{Istituto Nazionale di Fisica Nucleare (INFN), Laboratori Nazionali del Gran Sasso, 67100 Assergi, L’Aquila (AQ), Italy
}

\author{Valentina De Romeri~\orcidlink{0000-0003-3585-7437}}
\email{deromeri@ific.uv.es}
\affiliation{Instituto de F\'{i}sica Corpuscular (IFIC), CSIC‐Universitat de Val\'encia, E-46980 Valencia, Spain}

\author{Giulia Pagliaroli~\orcidlink{0000-0002-6751-9996}}
\email{giulia.pagliaroli@lngs.infn.it}
\affiliation{Istituto Nazionale di Fisica Nucleare (INFN), Laboratori Nazionali del Gran Sasso, 67100 Assergi, L’Aquila (AQ), Italy
}

\author{Dimitrios K. Papoulias~\orcidlink{0000-0003-0453-8492}}
\email{dimitrios.papoulias@uni-hamburg.de}
\affiliation{Institute of Experimental Physics, University of Hamburg, 22761, Hamburg, Germany}

\author{Federica Pompa~\orcidlink{0000-0002-9591-8361}}
\email{federica.pompa@lngs.infn.it}
\affiliation{Istituto Nazionale di Fisica Nucleare (INFN), Laboratori Nazionali del Gran Sasso, 67100 Assergi, L’Aquila (AQ), Italy
}

\author{Christoph A. Ternes~\orcidlink{0000-0002-7190-1581}}
\email{christoph.ternes@lngs.infn.it}
\affiliation{Gran Sasso Science Institute, Viale F. Crispi 7, L’Aquila, 67100, Italy}
\affiliation{Istituto Nazionale di Fisica Nucleare (INFN), Laboratori Nazionali del Gran Sasso, 67100 Assergi, L’Aquila (AQ), Italy
}

\keywords{neutrino invisible decay, dark matter detectors, solar neutrinos, \cevns, future sensitivity}

\begin{abstract}
The combination of the long baseline and characteristic energies of solar neutrinos offers an ideal framework to probe invisible neutrino  decay.
In this work we present the first constraint on invisible solar-neutrino decay using coherent elastic neutrino–nucleus scattering, recently observed in dark matter direct detection experiments. 
Through a combined analysis of nuclear-recoil data from XENONnT, PandaX-4T, and LUX-ZEPLIN, we constrain the lifetime of the neutrino mass state $\nu_2$, obtaining a bound already comparable in strength to that from the Sudbury Neutrino Observatory.
We further evaluate the sensitivity that could be reached by a future xenon-based dark matter detector. For this projection, we extend the analysis to electronic-recoil data, estimating the impact of detecting lower-energy solar neutrinos from the $pp$-chain via elastic scattering off electrons. 
This channel would allow us to place strong constraints on the lifetimes of both the $\nu_1$ and $\nu_2$ mass eigenstates. Our results show that, with nominal future exposures, nuclear-recoil data would improve the current bound by about one order of magnitude, while electronic-recoil data would open a new detection channel for low-energy solar neutrinos, surpassing existing dedicated solar-neutrino bounds by 1 to 2 orders of magnitude.

\end{abstract}
\maketitle

\section{Introduction}
Solar neutrinos have played a crucial role over the last 50 years. On one hand, they provided the first evidence of neutrino flavor conversion, resolving the solar neutrino problem~\cite{Davis:1968cp, Kamiokande-II:1989hkh, GALLEX:1992gcp, Abdurashitov:1996dp, Super-Kamiokande:1998zvz, SNO:2002tuh}. On the other hand, they help constrain the composition of the solar interior, through measurements of the neutrino fluxes produced in the $pp$ chain and the CNO cycle, which are key inputs to Standard Solar Models (SSMs)~\cite{Maltoni:2015kca, Villante:2020adi}.
Thanks to the wealth of available observations, the overall picture of solar neutrinos is now well established, although several non-standard phenomena remain viable as subleading effects.
One of these possibilities is the (non-standard) decay of neutrinos. The decay products may include one or more active neutrinos, or may consist exclusively of invisible particles. The former case is therefore referred to as \textit{visible decay}, while the latter one is referred to as \textit{invisible decay}.
In this work, we focus on invisible decay, which has been the subject of extensive study in the literature, see for example Refs.~\cite{Frieman:1987as,Lindner:2001fx,Beacom:2002cb,Gonzalez-Garcia:2008mgl,Gomes:2014yua,Berryman:2014qha,Pagliaroli:2015rca,Abrahao:2015rba,Abrahao:2015rba,Picoreti:2015ika,Choubey:2017dyu,Choubey:2017eyg,deSalas:2018kri,Denton:2018aml,Choubey:2018cfz,SNO:2018pvg,Ghoshal:2020hyo,Choubey:2020dhw,Pompa:2023yzg,Chakraborty:2020cfu,Ghoshal:2020hyo,Chattopadhyay:2021eba,Chattopadhyay:2022ftv,Ivanez-Ballesteros:2023lqa,Banerjee:2023sxj,KM3NeT:2023ncz,Valera:2024buc,Ternes:2024qui,Martinez-Mirave:2024hfd,Beccaria:2026ous}. The flux of neutrinos with mass state $i$ is suppressed by a factor
\begin{equation}
    \mathcal{D}_i(E_\nu, L)=\exp\left(-\alpha_i \frac{L}{E_\nu}\right)  \, ,
    \label{eq:damp_fac}
\end{equation}
that depends on the neutrino energy $E_\nu$, the distance between the source and the detector $L$, and the decay parameter
\begin{equation}
    \alpha_i=m_i/\tau_i \, ,
\end{equation}
where $m_i$ is the mass and $\tau_i$ the lifetime of the neutrino mass eigenstate $\nu_i$. These expressions show that the sensitivity to $\alpha_i$ depends on the ratio $L/E_\nu$, resulting in stronger sensitivities for large baselines and small energies. 
Currently, the constraints on $\alpha_3$ come from a joint analysis of MINOS/MINOS+~\cite{MINOS:2017cae}, NOvA~\cite{NOvA:2021nfi}, and T2K~\cite{T2K:2023smv} data ($\alpha_3\leq 2.7 \times10^{-5}$ eV$^2$ at 90\% confidence level (CL),~\cite{Ternes:2024qui}). Combining long-baseline accelerator experiments MINOS and K2K~\cite{K2K:2006yov} with atmospheric neutrino data from Super-Kamiokande~\cite{Super-Kamiokande:2006jvq} yields $\alpha_3\leq 2.3 \times10^{-6}$ eV$^2$ at 90\% CL~\cite{Gonzalez-Garcia:2008mgl}, driven by the the Super-Kamiokande data. 
The most stringent bound on $\alpha_1\sim\alpha_2$ comes from supernova SN1987A\footnote{Note, however, that SN1987A data cannot be used to place individual bounds on $\alpha_1$ and $\alpha_2$ in the case of normal neutrino mass ordering; see Fig.~1 in Ref.~\cite{Martinez-Mirave:2024hfd}.} ($\alpha_1\sim\alpha_2 \lesssim 1.2 \times 10^{-21}$ eV$^2$,~\cite{Frieman:1987as,Ivanez-Ballesteros:2023lqa, Martinez-Mirave:2024hfd}), while solar neutrino data have also been used to constrain $\alpha_1$ and $\alpha_2$ at 95\% CL, yielding $\alpha_1 \leq 1.6 \times 10^{-13}$ eV$^2$ and $\alpha_2 \leq 9.3 \times 10^{-13}$ eV$^2$~\cite{Berryman:2014qha}. Solar neutrinos have also been recognized as a promising probe of invisible neutrino decay because of their large propagation distance and relatively low energies, which maximize the ratio $L/E_\nu$.
In particular, Ref.~\cite{Martinez-Mirave:2024hfd} investigated the sensitivity of future experiments like DARWIN and RES-NOVA to invisible neutrino decay using solar and supernova neutrinos (see also Ref.~\cite{Huang:2018nxj}), showing that next-generation experiments could significantly improve existing bounds.

In this work we exploit, for the first time, the recent observation of solar-neutrino-induced \cevns~in dark matter direct detection experiments to search for invisible neutrino decay. Using the recently released XENONnT, PandaX-4T, and LZ data, we derive the first constraints on invisible neutrino decay from solar neutrinos detected through the \cevns\ channel. Thanks to their high exposures, extremely low-energy thresholds, and ultra-low backgrounds, dark matter direct detection experiments, have now entered the so-called \emph{neutrino fog}~\cite{OHare:2021utq,Monroe:2007xp,Vergados:2008jp,Strigari:2009bq,Billard:2013qya,DeRomeri:2025nkx} regime, where dark matter signals become difficult to disentangle from coherent elastic neutrino-nucleus scattering (\cevns)~\cite{Freedman:1973yd,Abdullah:2022zue}. The first evidence of {\cevns} induced by $^8$B solar neutrinos was reported by the XENONnT~\cite{XENON:2024ijk,XENON:2026ydt}, PandaX-4T~\cite{PandaX:2024muv}, and LUX-ZEPLIN (LZ)~\cite{LZ:2025igz} collaborations. These data have already been used to test the Standard Model (SM) and some possible extensions thereof, including nonstandard neutrino interactions~\cite{AristizabalSierra:2024nwf,Li:2024iij,Gehrlein:2025isp,DeRomeri:2026prc}, sterile neutrinos~\cite{Kelly:2026avh}, new light mediators~\cite{DeRomeri:2024iaw,Blanco-Mas:2024ale,Xia:2024ytb,AtzoriCorona:2025gyz}, and neutrino electromagnetic properties~\cite{DeRomeri:2024hvc}. 
Interestingly, they can be also used to constrain the decay parameter $\alpha_2$. 
Indeed, the nuclear-recoil threshold of current Xe-based detectors allows the detection of neutrinos with energies above approximately 7 MeV~\cite{XENON:2026ydt, PandaX:2024muv, LZ:2025igz}. At these energies, the Mikheyev–Smirnov–Wolfenstein (MSW) effect~\cite{Wolfenstein:1977ue, Mikheyev:1985zog} together with incoherent propagation between the Sun and the Earth causes that more than 90\% of all $^8$B neutrinos arrive as $\nu_2$; the experiments are therefore sensitive to $\alpha_2$ alone, with no sensitivity to $\alpha_1$ or $\alpha_3$.

The constraints on neutrino decay achievable with dark matter direct detection experiments can be expanded by future, larger detectors. 
In the second part of our analysis, we assess the impact of the next-generation xenon observatory XLZD~\cite{XLZD:2024nsu} under assumptions of nominal exposure, adopting scaling arguments for both signal and background. In this scenario, in addition to nuclear-recoil {\cevns} induced by solar $^8$B neutrinos, elastic scattering on electrons (\eves) induced by low-energy $pp$ neutrinos, at present facilities subdominant given current exposures and background levels, would also become detectable. 
This channel could provide further tests of the SM and its extensions, as shown by several sensitivity studies~\cite{Essig:2018tss,DARWIN:2020bnc,deGouvea:2021ymm,Mishra:2023jlq,Giunti:2023yha,DeRomeri:2024dbv,Celestino-Ramirez:2025snn,Gehrlein:2025isp,AtzoriCorona:2025xwr}, and, as we demonstrate here, in the context of neutrino decay, where it can also constrain $\alpha_1$, confirming earlier predictions discussed in~\cite{Martinez-Mirave:2024hfd,Huang:2018nxj}.
These complement the searches that can be performed with nuclear-recoil data at this observatory, see for example Refs.~\cite{Harnik:2012ni,Cerdeno:2016sfi,Dutta:2017nht,Gelmini:2018gqa,AristizabalSierra:2020zod,Amaral:2020tga,Dutta:2020che,Suliga:2020jfa,Amaral:2021rzw,Aalbers:2022dzr,Alonso-Gonzalez:2023tgm,Amaral:2023tbs}.

The remainder of this paper is organized as follows. In Sec.~\ref{sec:decay} we discuss how invisible neutrino decay modifies the conversion probabilities of solar neutrinos. In Sec.~\ref{sec:stat} we outline our procedure for computing event rates and our statistical analysis. In Sec.~\ref{sec:res} we present our results, and in Sec.~\ref{sec:conc} we conclude.


\section{Invisible neutrino decay}
\label{sec:decay}

\begin{figure}[t!]
    \centering
    \includegraphics[width=\textwidth]{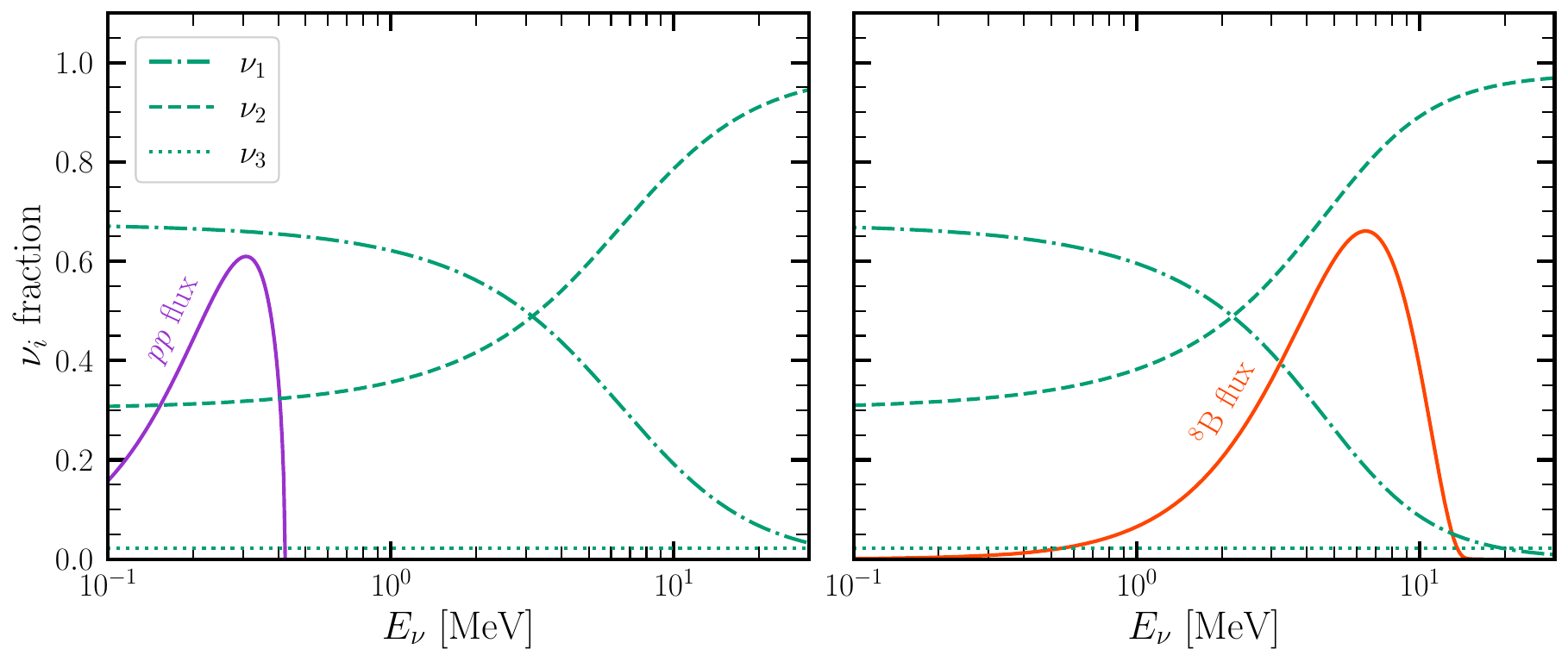}
    \caption{Fraction of each neutrino mass eigenstate as a function of energy, assuming full adiabatic evolution, for the $pp$ (\textbf{left panel}) and $^8$B (\textbf{right panel}) neutrino production densities. The corresponding $pp$ and $^8$B fluxes are also shown, in arbitrary units, to indicate the energy range relevant to each channel, \eves~and \cevns}.
    \label{fig:frac}
\end{figure}

In the presence of invisible neutrino decay the neutrino flavor evolution is described by the Hamiltonian 
\begin{equation}
 H = \frac{1}{2E_\nu} \left[H_0 + H_m + H_D\right],
 \label{Ham_decay}
\end{equation}
where the first two terms correspond to the standard  vacuum and  matter terms, 
\begin{eqnarray}
H_0 &=& U\diag(0,\Delta m_{21}^2,\Delta m_{31}^2) U^\dagger\,,\\
H_m &=& 
\diag(V,0,0)\,,
\end{eqnarray}
where $U$ is the PMNS matrix, $V=2E_\nu\sqrt{2}G_F n_e$,  $G_F$ the Fermi constant and $n_e$ the electron number density. The last term in Eq.\ (\ref{Ham_decay}) represents the neutrino decay contribution
\begin{equation}
 H_D = U\diag(-i\alpha_1,-i\alpha_2, -i\alpha_3)
U^\dagger .
\end{equation}
Since the distance traveled by neutrinos between the Sun and Earth is much larger than the distance from their production point to the solar surface, it is safe to assume standard evolution inside the Sun, with the decay occurring only along the Sun-Earth path. Under this assumption, the flavor conversion probability is given by
\begin{equation}
    \mathcal{P}_{e \alpha}(E_\nu,r) = \sum_{i=1}^3 |U_{e i}^m(E_\nu,r)|^2 |U_{\alpha i}|^2 \mathcal{D}_i(E_\nu, L_\odot) \,,
    \label{eq.full_prob}
\end{equation}
where $r$ is the neutrino production point in the Sun and $L_\odot = 150\times 10^6$~km is the Sun-Earth distance. The observable oscillation probability is obtained by integrating this expression over $r$, weighted by the production probability density, $\rho_X(r)$, of the relevant solar neutrino source,

\begin{equation}
    P^X_{e\alpha}(E_\nu) = \int \d r~ \rho_X(r)~\mathcal{P}_{e \alpha}(E_\nu,r)\,,
    \label{eq:rho_integral_prob}
\end{equation}
where, as already mentioned in the introduction, we will be mainly interested in $X=pp$ and $X={}^8$B. These densities are taken from the standard solar model named ``MB22m"~\cite{Magg:2022rxb,herrera_2023_10822316}.

Due to the MSW effect, the neutrino composition changes with energy. 
Assuming full adiabatic evolution, we show in Fig.~\ref{fig:frac} the fraction of each mass state as a function of energy. These fractions are obtained by weighting $|U_{e i}^m(E_\nu,r)|^2$ with the production densities $\rho_{pp}(r)$ (left panel) and $\rho_{{}^8\textrm{B}}(r)$ (right panel). 
Here, we are mainly interested in very high-energy neutrinos ($E_\nu\gtrsim 7$~MeV, relevant for \cevns) and very low-energy neutrinos ($E_\nu\lesssim0.4$~MeV, relevant for \eves) regimes of the solar flux. 
The fluxes relevant at these energies come from $^8$B neutrinos and from $pp$ neutrinos, respectively, which we also report in Fig.~\ref{fig:frac} in arbitrary units. 
Subdominant contributions from the $hep$ and $^7$Be components can also contribute to the \cevns~and \eves~rates, respectively, albeit at a negligible level. As can be seen, in the high-energy, matter-dominated, regime the flux is dominated by $\nu_2$, so we do not expect any sensitivity to $\nu_1$ or $\nu_3$. In the low-energy regime, the flux is dominated instead by $\nu_1$. We do not expect any meaningful sensitivity from solar neutrinos to $\alpha_3$, since the fraction of $\nu_3$ of the total flux is always very small.

\begin{figure}[t!]
    \centering
    \includegraphics[width=\textwidth]{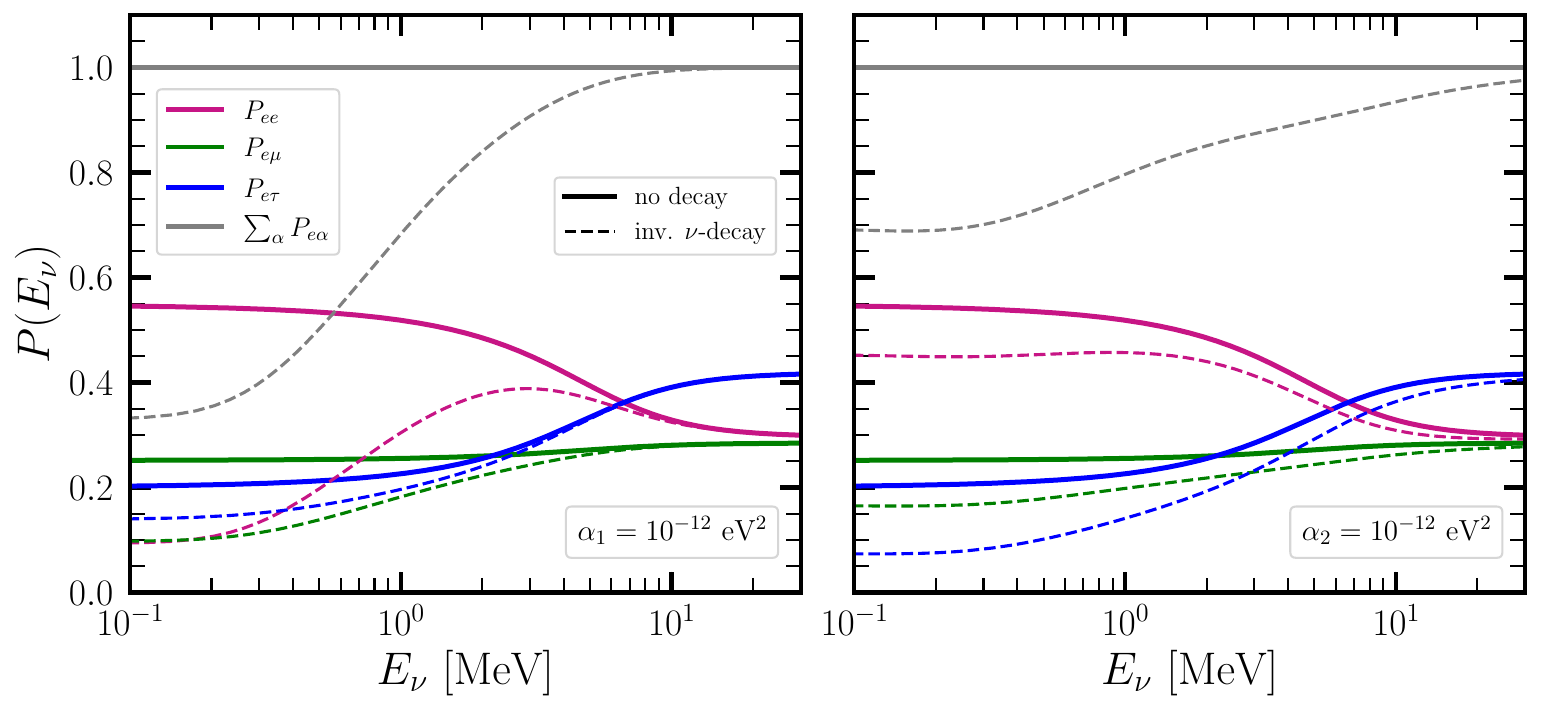}
    \caption{Neutrino flavor conversion probabilities as a function of the neutrino energy in presence of neutrino decay (dashed lines), for representative values of $\alpha_1 = 10^{-12}$ eV$^2$ (\textbf{left panel}) and $\alpha_2 = 10^{-12}$ eV$^2$ (\textbf{right panel}). The same probabilities in the absence of invisible neutrino decay are also displayed for comparison (solid lines).}
    \label{fig:probs}
\end{figure}

To gain analytical insight, we derive approximate expressions for both energy regimes; our numerical calculations, however, always rely on the full expression in Eq.~(\ref{eq.full_prob}). For the high-energy $^8$B neutrinos, we are in the  scenario where the effective mixing angle in matter is $\sin^2\theta_{12}^m = 1$. In this case, $|U_{e1}^m|^2 = 0$, $|U_{e2}^m|^2 = \cos^2\theta_{13}$, and $|U_{e3}^m|^2 = \sin^2\theta_{13}$. Using also the fact that {\cevns} is flavor-blind, the effective probability becomes
\begin{equation}
    P^\textrm{HE}(E_\nu) = \sum_\alpha P^{^8\textrm{B}}_{e\alpha}(E_\nu) = \cos^2\theta_{13} \exp\left(-\alpha_2 \frac{L_\odot}{E_\nu} \right) + \sin^2\theta_{13} \exp\left(-\alpha_3 \frac{L_\odot}{E_\nu} \right)\,.
\end{equation}
On the other hand, for the very low-energy $pp$ neutrinos, we are in the vacuum regime, where $|U_{e i}^m(E_\nu,r)|^2 = |U_{e i}|^2$. The oscillation probabilities therefore reduce to
\begin{equation}
    P^{\textrm{LE}}_{e \alpha}(E_\nu) = \sum_{i=1}^3 |U_{e i}|^2 |U_{\alpha i}|^2 \exp\left(-\alpha_i \frac{L_\odot}{E_\nu} \right)\,.
\end{equation}
In our calculations, we fix the mixing angles to $\sin^2\theta_{12} = 0.309$, $\sin^2\theta_{13} = 0.022$, $\sin^2\theta_{23} = 0.5$~\cite{deSalas:2020pgw,JUNO:2025gmd} (see also Refs.~\cite{Esteban:2024eli,Capozzi:2025wyn}), and the mass splitting to $\Delta m_{21}^2=7.5\times 10^{-5}$~eV$^2$, setting $\delta = 0$. 
We have verified that varying these parameters within their allowed ranges leads to negligible changes in our results. 
In Fig.~\ref{fig:probs} we show representative examples of how the decay of $\nu_1$ (left panel) or $\nu_2$ (right panel) affects neutrino oscillation probabilities (dashed lines versus the solid ones, which depict the SM scenario). 
The latter have been calculated using Eq.~\eqref{eq:rho_integral_prob} with $X={}^8$B. 
The probabilities corresponding to $pp$ neutrinos are very similar, and a comparison is provided in Appendix~\ref{sec:appA}.
As shown in the left panel of Fig.~\ref{fig:probs}, the decay of $\nu_1$ has little impact on high-energy neutrinos but leads to a strong suppression at low energies. The effect of $\nu_2$ decay, shown in the right panel, is smaller in magnitude but visible across all relevant energies.


\section{Simulation and analysis details}
\label{sec:stat}

Neutrinos can interact via \cevns~with the xenon nuclei in the detector. The cross section for this process is given by~\cite{Freedman:1973yd,Barranco:2005yy}
\begin{equation}
\label{eq:xsec_CEvNS_SM}
\frac{\d\sigma_{\nu_\ell \mathcal{N}}}{\d T_{\mathcal N}}
= \frac{G_F^2\, m_{\mathcal N}}{\pi}
\left(Q_{V,\ell}^{\rm SM}\right)^2
F_W^2(\qtransfer^2)
\left(1 - \frac{m_{\mathcal N} T_{\mathcal N}}{2 E_\nu^2}\right),
\end{equation}
where $m_{\mathcal N}$ is the nuclear mass, $T_{\mathcal N}$ the nuclear-recoil energy and $\qtransfer = \sqrt{2 m_{\mathcal N} T_{\mathcal N}}$ the momentum transfer. The quantity $Q_{V,\ell}^{\rm SM}$ is the SM weak vector charge of the nucleus,
\begin{equation}
\label{eq:CEvNS_SM_QV}
Q_{V,\ell}^{\rm SM} = g_{V,\ell}^p\, Z + g_{V,\ell}^n\, N ,
\end{equation}
where $Z$ and $N$ are the proton and neutron numbers, respectively. It depends on the vector couplings for protons, $g_{V,\ell}^p = (1/2 - 2 \sin^2\theta_W)/2$, and neutrons, $g_{V,\ell}^n = -1/2$. The flavor dependence of these quantities becomes relevant only beyond tree level; see, e.g., Ref.~\cite{Cadeddu:2020lky}.
The weak charge depends on the weak mixing angle, whose value at low energies is $\sin^2\theta_W = 0.23857(5)$~\cite{ParticleDataGroup:2024cfk}.
The weak nuclear form factor $F_W(\qtransfer^2)$ accounts for the finite spatial extent of the nucleus. For the energies relevant in the experiments considered here, the typical momentum transfer is relatively small resulting in only a modest suppression. We adopt the Klein-Nystrand parametrization~\cite{Klein:1999qj}, given by
\begin{equation}
\label{eq:KNFF}
F_W(\qtransfer^2)
= 3\,\frac{j_1(\qtransfer R_A)}{\qtransfer R_A}
  \left(\frac{1}{1 + \qtransfer^2 a_k^2}\right),
\end{equation}
where $j_1(x) = \sin x/x^2 - \cos x/x$ is the spherical Bessel function of first order, $a_k = 0.7~\mathrm{fm}$, and $R_A = 1.23\,A^{1/3}$ (in fm) denotes the nuclear root-mean-square radius, with $A$ the atomic mass number. 

Another possibility to detect solar neutrinos at dark matter direct detection experiments is via the \eves~channel. The differential \eves~cross section with respect to the electron-recoil energy $T_e$ is given by
\begin{align}
\label{eq:xsec_vES_SM}
\begin{split}
\frac{\d\sigma_{\nu_{\ell} \mathcal{A}}}{\d T_e}=& Z^\mathcal{A}_\mathrm{eff}(T_e) \frac{G_F^2m_e}{2\pi}\left[(g_{V,\ell}^e + g_{A,\ell}^e)^2 +(g_{V,\ell}^e - g_{A,\ell}^e)^2\left(1-\frac{T_e}{E_\nu}\right)^2 -((g_{V,\ell}^e)^2-(g_{A,\ell}^e)^2)\frac{m_e T_e}{E_\nu^2}\right] \, ,
\end{split}
\end{align}
where $m_e$ denotes the electron mass. The vector and axial-vector couplings depend on the flavor $\ell$ of the incoming neutrino and their tree-level values are given by $g_{V,\ell}^e=2 \sin^2 \theta_W - 1/2 + \delta_{{\ell} e}$, $g_{A,\ell}^e=-1/2 + \delta_{{\ell} e}$. The term $\delta_{{\ell} e}$ arises because, for an incoming electron neutrino, both neutral- and charged-current interactions contribute, while for $\nu_{\mu}$ and $\nu_{\tau}$, \eves~receives contributions from neutral currents only. Assuming that the target electrons are bound in the atoms $\mathcal{A}$ of the detector material, the factor $Z^\mathcal{A}_\text{eff}(T_e)$ accounts for the effective number of electrons that can be ionized at a given energy $T_e$. Here we take the relevant values for xenon from the Hartree-Fock calculations provided in Ref.~\cite{Chen:2016eab}. 

The differential event rate at the detector $d$ is obtained by convolving the oscillated neutrino flux 
\begin{equation}
    \frac{\d\phi_\ell}{\d E_\nu} = \frac{\d\phi_e^0}{\d E_\nu} P_{e\ell} \, ,
\end{equation}
where $\frac{\d\phi_e^0}{\d E_\nu}$ is the unoscillated electron-neutrino flux, with the interaction cross section. For the unoscillated flux we use the prediction of the MB22m standard solar model~\cite{Magg:2022rxb,herrera_2023_10822316}. The integrated fluxes of $pp$ and $^8$B neutrinos in this model are
\begin{eqnarray}
    \Phi_{pp} &=& 5.950 ~(1\pm0.0059)\times 10^{10} \frac{1}{\textrm{s cm}^2}\,,
    \\
    \Phi_{{}^8\textrm{B}} &=& 5.127~ (1\pm0.1313)\times 10^{6} \frac{1}{\textrm{s cm}^2}\,.
\end{eqnarray}
The rate is then given by
\begin{equation}
    \frac{\d R^d}{\d T} =  \mathcal{E}^d \mathcal{F}^d(T) \int_{E_\nu^\text{min}}^{E_\nu^\text{max}} \d E_\nu \sum_{\ell=e,\mu,\tau}\frac{\d\phi_\ell}{\d E_\nu} \frac{\d\sigma_{\nu_\ell}}{\d T}\,,
    \label{eq:diff_rate}
\end{equation}
where $\frac{\d\sigma_{\nu_\ell}}{\d T}$ is either the \cevns~or the \eves~cross section. In the former case, $T$ refers to the nuclear-recoil energy $T_{\mathcal N}$, while in the latter it refers to the electron-recoil energy $T_e$. 
As stated above, for nuclear recoils the dominant contribution to the solar neutrino flux comes from $^8$B neutrinos, while for electron recoils it comes from $pp$ neutrinos. 
We have verified that the bounds obtained in this paper are unchanged if the other flux components are included, given their subdominant contribution. 
In Eq.~\eqref{eq:diff_rate}, $\mathcal{E}^d$ is the exposure of detector $d$, and $\mathcal{F}^d(T)$ is its energy-dependent efficiency. 
The rate in Eq.~\eqref{eq:diff_rate}, differential in recoil energy, still needs to be translated into a rate in the observed variable. 
The following subsections briefly describe how this is done for each experiment. Note that neutrino decay modifies the observable event spectrum exclusively through the oscillated neutrino flux, while the interaction cross sections (unlike in many other possible scenarios of physics beyond the SM studied with \cevns) remain unchanged.

\subsection{PandaX-4T}

Our analysis follows Refs.~\cite{DeRomeri:2024iaw,DeRomeri:2024hvc}. The data in Ref.~\cite{PandaX:2024muv} are presented in bins of number of electrons, $N_{e^-}$. The events per bin are then given by
\begin{equation}
    R_i^\mathrm{P} = c_i\mathcal{E}^{\mathrm{P}}\int_i \dfrac{\d R^\mathrm{P}}{\d n^\mathrm{P}} \d n^\mathrm{P}\,,
\label{eq:ev_rate}    
\end{equation}
where the integral is performed over the size of bin $i$, while $n^\text{P}=N_{e^-}$. We consider 8 bins in the range [4, 8] $N_{e^-}$, corresponding to nuclear-recoil energies in the range $[0.66, 1.19]$~keV. 
The correction factors $c_i$ in Eq.~\eqref{eq:ev_rate} are included to match our predictions with the best-fit spectra presented in Ref.~\cite{PandaX:2024muv}. 
These factors can be regarded as effective efficiencies, included because our analysis is simplified relative to that of the PandaX collaboration: we use only S2 information, whereas the experimental analysis relies on several additional variables fitted simultaneously in a correlated way. 
The differential event rate is expressed via a change of variables,
\begin{equation}
    \dfrac{\d R^\mathrm{P}}{\d n^\mathrm{P}} = \dfrac{\d R^\mathrm{P}}{\d T_\mathcal{N}} \dfrac{\d T_\mathcal{N}}{\d n^\mathrm{P}}\, ,
    \label{eq:T2n_varchange}
\end{equation}
where
\begin{equation}
    n^\mathrm{P} = N_{e^-} = T_\mathcal{N}Q_y^\mathrm{P}(T_\mathcal{N})\, ,
\end{equation}
with the charge yield $Q_y^\mathrm{P}(T_\mathcal{N})$ given in Ref.~\cite{PandaX:2024muv}. 

\subsection{XENONnT}

We use the full XENONnT dataset collected during the three science runs (SR0, SR1, SR2), corresponding to a total exposure of 6.77~$\mathrm{t \times yr}$~\cite{XENON:2026ydt}. Our analysis follows the procedure described in Ref.~\cite{DeRomeri:2026prc}. In particular, we analyze each SR individually, using all observables provided in the data release: the corrected secondary scintillation signal (cS2), the ratio $\mathrm{S2}/\Delta t$ (where $\Delta t$ is the drift time between the S1 and S2 signals and encodes the interaction depth), and the boosted decision tree (BDT) scores for the S1 and S2 signals, which quantify the compatibility of an event waveform with a genuine recoil. These observables are combined in a simultaneous four-dimensional (4D) binned analysis with 81 bins per SR, for a total of 243 bins. The \cevns~signal is computed assuming SR-specific exposures of (1.174, 2.343, 3.250)~$\mathrm{t \times yr}$ for (SR0, SR1, SR2), respectively. Detector response effects are modeled using the SR-dependent 4D response matrices provided in the XENONnT \cevns~data release~\cite{XENONnT_cevns_data_release,XENONnT_github}, which map the true nuclear-recoil energy onto the four reconstructed observables. Each observable is divided into three bins following the XENONnT binning scheme, and the response matrices encode the probability for a recoil of a given energy to populate any combination of these bins. Finally, the 4D templates describing the accidental-coincidence (AC), electron-recoil, and neutron-recoil backgrounds are also taken from the XENONnT \cevns~data release~\cite{XENONnT_cevns_data_release}.

\subsection{LUX-ZEPLIN}

Our analysis of the LZ data follows Ref.~\cite{DeRomeri:2026prc}.
The data are presented in bins of S2~\cite{LZ:2025igz}. As above, the number of events per bin $i$ is given by
\begin{equation}
    R^\textrm{LZ}_i = \int_i \frac{\d R^\textrm{LZ}}{\d n^\textrm{LZ}} \d n^\textrm{LZ}\,.
    \label{eq:diff_rate_n}
\end{equation}
Within the experimental region of interest (ROI) and following \cite{LZ:2025igz}, we consider $5$ bins between $[155, 645]$ photons detected (phd) corresponding to nuclear-recoil energies in the range $[1, 6]$~keV.
The differential event rate in Eq.~(\ref{eq:diff_rate_n}) is then expressed via a change of variables
\begin{equation}
    \frac{\d R^\textrm{LZ}}{\d n^\textrm{LZ}} = \frac{\d R^\textrm{LZ}}{\d T_{\mathcal{N}}} \frac{\d T_{\mathcal{N}}}{\d n^\textrm{LZ}}\,,
    \label{eq:change_variables}
\end{equation}
where the mapping between the nuclear-recoil energy and the S2 signal is
\begin{equation}
    n^\textrm{LZ} \equiv S2 = g_2T_{\mathcal{N}} Q_y(T_{\mathcal{N}})\,,
    \label{eq:nS2}
\end{equation}
with $g_2 = 34.0$ phd/electron, and the charge yield $Q_y(T_{\mathcal{N}})$ given in Ref.~\cite{LZ:2025igz}.

\subsection{XLZD}

We also compute the sensitivity of the future XLZD observatory to invisible neutrino decay, considering both nuclear and electron recoils. For nuclear recoils, our simulation is based on the LZ analysis described above but scaled-up to exposures of 200 and 1000~$\mathrm{t \times yr}$. Since the final detector design has not yet been established, we will consider an optimistic and a pessimistic scenario, in which the background components are scaled by an additional factor of 0.5 and 2, respectively.

For electron recoils, our analysis follows Refs.~\cite{Giunti:2023yha,DeRomeri:2024dbv} (referred to as DARWIN therein). In this case, the number of events in bin $i$ is obtained from
\begin{equation}
   R^{\rm E\nu ES}_i = \int_{T_e^i}^{T_e^{i+1}}\d T_e ~\int_0^{T_{e'}^\mathrm{max}} \d T_e'~ R(T_e,T_e')~ \frac{\d R}{\d T_e'}\,,
\label{eq:r_eves}
\end{equation}
where $\frac{\d  R}{\d T_e'}$ is given in Eq.~\eqref{eq:diff_rate}, $T_{e'}^{\mathrm{max}}$ denotes the maximum true electron recoil energy obtained for neutrinos at the endpoint of the $pp$ neutrino spectrum, while $T_e^i$ stands for the reconstructed electron recoil energy in the $i$th bin. 
In the absence of more detailed information, we use the resolution function $R(T_e,T_e')$ and detector efficiency reported by the XENONnT collaboration in Ref.~\cite{XENON:2022ltv}.

\subsection{Statistical analysis}
For each detector, the total number of predicted events in each bin $i$ is given by
\begin{equation}
    N_i = R_i(1+\gamma) + \sum_i B_i(1+\beta_i)\,,
    \label{eq:total_in_bin}
\end{equation}
where the spectrum of the dominant backgrounds $B_i$ is provided by each Collaboration \cite{XENON:2026ydt,PandaX:2024muv,LZ:2025igz,DARWIN:2020bnc}. We compare our predictions with the measured data $D_i$ (in the case of XLZD we generate a fake data set without neutrino decay) using
\begin{equation}
    \chi^2 = \min\limits_{\gamma,\beta_k} \Bigg\{ 2 \Bigg[ \sum_i N_i - D_i + D_i  ~\text{ln}\Bigg( \frac{D_i}{N_i} \Bigg) \Bigg] + \Bigg( \frac{\gamma}{\sigma_{\gamma}} \Bigg)^2 + \sum_k\Bigg( \frac{\beta_k}{\sigma_{\beta_k}} \Bigg)^2  \Bigg\}\,.
    \label{eq:chi2_analysis}
\end{equation}
Here, $\gamma$ and $\beta_k$ are the nuisance parameters accounting for the neutrino flux predictions (with uncertainties~\cite{Magg:2022rxb,herrera_2023_10822316} $\sigma_\textrm{B} = 13.13\%$ and $\sigma_{pp} = 0.59\%$ for $^8$B and $pp$ neutrinos, respectively) and for the normalization of each background component (with associated uncertainties $\sigma_{\beta_k}$).

The XENONnT likelihood is constructed from a 4D $\chi^2$ analysis with 81 bins per SR (243 bins in total), as described above. Separate nuisance parameters are introduced for the three AC background components, one for each SR, while common nuisance parameters are used for the electron-recoil and neutron-background uncertainties across all SRs, following Ref.~\cite{XENON:2026ydt}. For the \cevns~signal, we include the uncertainty on the $^8$B solar neutrino flux normalization together with a 5\% uncertainty on the detector fiducial volume.


\section{Results}
\label{sec:res}

In this section we discuss the results of our analyses. We present the bounds obtained from the data of current experiments in Sec.~\ref{sec:res_current}, while Secs.~\ref{sec:res_future_NR} and~\ref{sec:res_future_ER} discuss the XLZD sensitivities achievable through nuclear- and electron-recoil measurements, respectively.

\subsection{Current bounds from solar \cevns~measurements}
\label{sec:res_current}

\begin{figure}[t!]
    \centering
    \includegraphics[width=0.49 \textwidth]{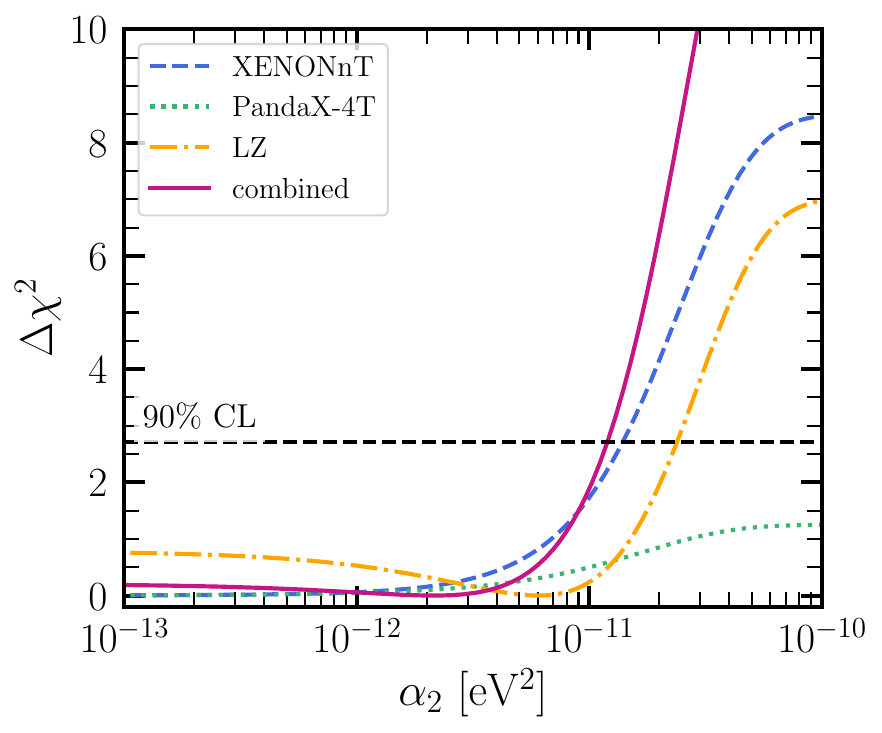}
    \includegraphics[width=0.49 \textwidth]{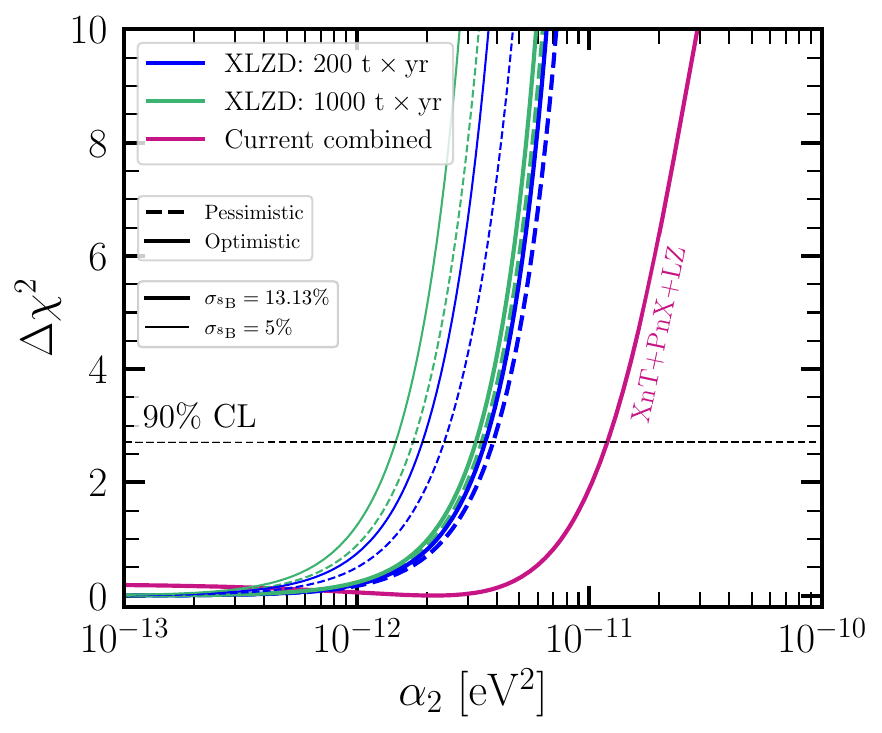}   
    \caption{$\Delta\chi^2$ profiles for $\alpha_2$ from \cevns~measurements. \textbf{Left panel:} current bounds from PandaX-4T (green, dotted), XENONnT (blue, dashed), and LZ (orange, dot-dashed), together with their combination (magenta). \textbf{Right panel:} projected sensitivity of the future XLZD experiment, for exposures of 200~$\mathrm{t \times yr}$ (blue) and 1000~$\mathrm{t \times yr}$ (green), under pessimistic (dashed) and optimistic (solid) background assumptions. Thick and thin lines correspond, respectively, to the current uncertainty on the $^8$B flux normalization ($\sigma_{^8\mathrm{B}}=13.13\%$) and to an improved future uncertainty ($\sigma_{^8\mathrm{B}}=5\%$).}
    \label{fig:CEvNS_bounds}
\end{figure}

The results of our analysis of current experimental data are shown in the left panel of Fig.~\ref{fig:CEvNS_bounds}, where we report the results from PandaX-4T (green, dotted), LZ (orange, dash-dotted), and XENONnT (blue, dashed), together with the combined analysis of all three experiments (magenta, solid).
We cannot place any significant bound on neutrino decay from PandaX-4T data alone: since we use a reduced version of the PandaX-4T dataset\footnote{As stated in Ref.~\cite{DeRomeri:2024iaw}, we have nonetheless verified that, in the SM analysis, our predicted number of events lies within the $1\sigma$ interval reported by the collaboration for the US2-only analysis~\cite{PandaX:2024muv}.} (only US2), we obtain a weaker constraint on the solar neutrino flux than the collaboration~\cite{DeRomeri:2024iaw}, which in turn translates into a weak bound on $\alpha_2$.
Using LZ and XENONnT data we are, on the other hand, able to place meaningful bounds on $\alpha_2$. 
Although the statistical significance of the \cevns~observation is higher in LZ than in XENONnT, we find that XENONnT yields the stronger bound. 
This is because LZ has measured slightly fewer events than expected under the SM hypothesis, which results in a nonzero best-fit value for $\alpha_2$ and a correspondingly shifted $\chi^2$ profile. We show in Appendix~\ref{sec:appB} that the expected sensitivity of LZ -- i.e., the sensitivity obtained assuming the data exactly match the SM \cevns~prediction -- is, in fact, stronger than the observed exclusion limit.
The combined profile, also shown in magenta in both panels of Fig.~\ref{fig:CEvNS_bounds}, yields a bound of
\begin{equation}
 \alpha_2 < 1.2 \times10^{-11}~\textrm{eV}^2\,,
\end{equation}
at 90\% CL. Nevertheless relying only on the recently observed solar \cevns~signal, this bound is already comparable in strength to the one derived from the combined analysis of the three phases of SNO data~\cite{SNO:2018pvg}, $\alpha_2<8.1\times10^{-12}$~eV$^2$, despite being obtained through a completely independent detection channel and with only a fraction of the total exposure eventually expected from dark matter direct detection experiments. This result highlights the strong potential of \cevns-based searches to become competitive with dedicated solar-neutrino experiments as exposures grow.
It is worth stressing that solar neutrinos are intrinsically better suited to probe $\alpha_2$ than neutrinos from terrestrial sources. As evident from Eq.~\eqref{eq:damp_fac}, the decay-induced damping becomes more pronounced with increasing baseline $L$. Since the Sun--Earth distance exceeds the baselines of atmospheric, accelerator, and reactor neutrino experiments by many orders of magnitude, solar neutrinos provide substantially greater sensitivity to $\alpha_i$.
Correspondingly, bounds on $\alpha_2$ from such terrestrial experiments are weaker: for instance, an analysis of JUNO data yields $\alpha_2 < 3\times10^{-6}$~eV$^2$ at 90\% CL~\cite{Beccaria:2026ous}, five orders of magnitude weaker than the bound obtained here. Solar neutrinos therefore remain the leading~\cite{Berryman:2014qha,Picoreti:2015ika} probes of $\alpha_2$, and are expected to stay so until next-generation solar, supernova or high-energy astrophysical neutrino observatories come online~\cite{Martinez-Mirave:2024hfd,Valera:2024buc}.

\subsection{Future bounds from solar \cevns~measurements}
\label{sec:res_future_NR}

We next discuss the sensitivity that can be achieved at the future XLZD facility via the \cevns~channel. The result of our analysis is shown in the right panel of Fig.~\ref{fig:CEvNS_bounds}. Dashed and solid lines correspond, respectively, to the pessimistic and optimistic background assumptions described in Sec.~\ref{sec:stat}.
Different colors represent different exposures (200~$\mathrm{t \times yr}$, blue, and 1000~$\mathrm{t \times yr}$, green), while thick and thin lines are obtained assuming, respectively, the current uncertainty on the solar $^8$B flux normalization ($\sigma_{^8\mathrm{B}}=13.13\%$) and an improved future uncertainty ($\sigma_{^8\mathrm{B}}=5\%$).
As the figure shows, the theoretical uncertainty on the $^8$B flux becomes the limiting factor for future experiments: increasing the exposure by a factor of five has almost no effect on the sensitivity as long as $\sigma_{^8\mathrm{B}}=13.13\%$ is retained. If this uncertainty is instead reduced, the bound strengthens considerably, as can be seen by comparing the thick and thin lines. This behavior reflects the fact that, once statistical uncertainties become subdominant, the sensitivity to $\alpha_2$ is limited by how precisely the unoscillated $^8$B flux can be predicted.
Future improvements in \cevns-based decay searches will depend at least as much on advances in solar modeling as on increased detector exposure.

Under the most optimistic assumptions (1000~$\mathrm{t \times yr}$ exposure, optimistic background scenario, and improved flux uncertainty), we find that a future observatory like XLZD could improve the current sensitivity by approximately one order of magnitude, reaching the $\mathcal{O}(10^{-12}~\mathrm{eV^2})$ level for $\alpha_2$. This would keep \cevns-based searches at dark matter direct detection experiments competitive with the projected reach of next-generation solar neutrino experiments~\cite{Martinez-Mirave:2024hfd}.

\subsection{Future bounds from solar \eves~measurements}
\label{sec:res_future_ER}

\begin{figure}[t!]
    \centering
    \includegraphics[width=0.49 \textwidth]{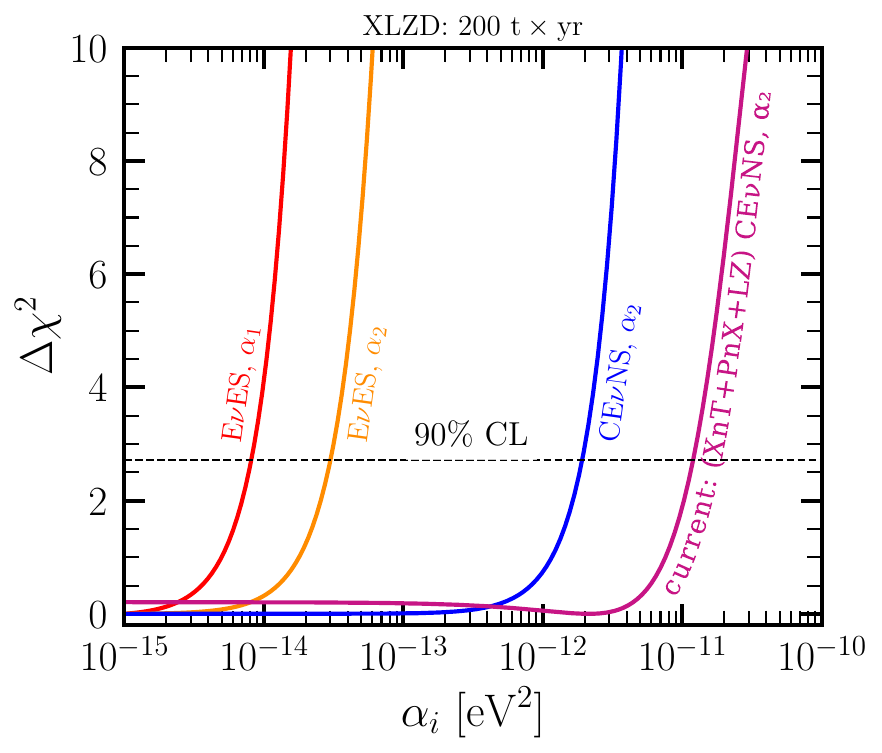}    
    \caption{Projected $\Delta\chi^2$ profiles for $\alpha_1$ (red) and $\alpha_2$ (orange) from the electron-recoil spectrum at the future XLZD experiment. For comparison, the projected sensitivity to $\alpha_2$ via the \cevns~channel at the same facility is also shown (blue), together with the current combined limit from XENONnT, PandaX-4T, and LZ (magenta).}
    \label{fig:future_EvES}
\end{figure}

Finally, we discuss the sensitivity that can be obtained through the measurement of $pp$ neutrinos via \eves. We do not consider currently running experiments in this channel, since only the PandaX-4T Collaboration has so far reported a slight indication of $pp$-neutrino detection, at approximately $2\sigma$ significance~\cite{PandaX:2026qdp}. 
The exposures and background levels of currently available data sets from dark matter experiments limit the detectability of $pp$ neutrinos via \eves, unless new physics that enhances the signal beyond the SM expectation is present (see, e.g., Ref.~\cite{DeRomeri:2024dbv}).
In future experiments (or with improved analyses of current data, as in Ref.~\cite{PandaX:2026qdp}) better control of background components is expected to enable a much more precise measurement of the $pp$ flux~\cite{DARWIN:2020bnc,XLZD:2024nsu}, in turn opening up sensitivity to neutrino decay in this channel.

As discussed in Sec.~\ref{sec:decay}, $pp$ neutrinos arrive at Earth mostly as $\nu_1$ and $\nu_2$, enabling the possibility to constrain $\alpha_1$ in addition to $\alpha_2$, unlike the \cevns~channel, which is sensitive to $\alpha_2$ alone. The projected sensitivities are shown in Fig.~\ref{fig:future_EvES}. As the figure shows, these bounds are considerably stronger than their \cevns~counterparts, for two main reasons. First, XLZD is expected to record a very large number of \eves~events, which would lead to an excellent measurement of the $pp$ flux. Second, $pp$ neutrinos have much lower energies than $^8$B neutrinos, which further enhance the sensitivity, since the decay terms scale as $E_\nu^{-1}$. We find that XLZD could constrain
\begin{eqnarray}
   \alpha_1 &<&  7.9 \times 10^{-15}~\textrm{eV}^2\,,\\
    \alpha_2 &<& 3.0 \times 10^{-14}~\textrm{eV}^2\,,
\end{eqnarray}
at 90\% CL.
The projected XLZD sensitivity therefore improves upon the current bounds from dedicated solar neutrino experiments~\cite{Berryman:2014qha,Picoreti:2015ika,SNO:2018pvg} by one to two orders of magnitude, demonstrating that electron-recoil measurements of $pp$ solar neutrinos in dark matter detectors provide one of the most sensitive probes of invisible neutrino decay. These projected sensitivities are consistent with the expectations presented in Refs.~\cite{Martinez-Mirave:2024hfd,Huang:2018nxj}, with small differences arising from the different assumptions adopted for the $pp$ solar neutrino flux uncertainty and the detector backgrounds.


\section{Conclusions}
\label{sec:conc}

\begin{table}[t]
\centering
\begin{tabular}{lcc}
\hline\hline
  & Parameter~~ & ~~Bound (90\% CL) [eV$^2$] \\
\hline
\multicolumn{3}{c}{\textbf{current experiments, \cevns}} \\
\hline
XENONnT & $\alpha_2$ & $1.40\times10^{-11}$ \\
PandaX-4T & $\alpha_2$ & --- \\
LZ & $\alpha_2$ & $2.41\times10^{-11}$ \\
Combined & $\alpha_2$ & $1.20\times10^{-11}$ \\
\hline
\multicolumn{3}{c}{\textbf{XLZD, \cevns}} \\
\hline
200~~$\mathrm{t \times yr}$ (optimistic) & $\alpha_2$ & $3.51\times10^{-12}$ \\
1000~$\mathrm{t \times yr}$ (optimistic) & $\alpha_2$ & $3.25\times10^{-12}$ \\
200~~$\mathrm{t \times yr}$ (pessimistic) & $\alpha_2$ & $3.94\times10^{-12}$ \\
1000~$\mathrm{t \times yr}$ (pessimistic) & $\alpha_2$ & $3.51\times10^{-12}$ \\
\hline
200~~$\mathrm{t \times yr}$ (optimistic, reduced $^8$B uncertainty) & $\alpha_2$ & $1.90\times10^{-12}$ \\
1000~$\mathrm{t \times yr}$ (optimistic, reduced $^8$B uncertainty) & $\alpha_2$ & $1.45\times10^{-12}$ \\
200~~$\mathrm{t \times yr}$ (pessimistic, reduced $^8$B uncertainty) & $\alpha_2$ & $2.39\times10^{-12}$ \\
1000~$\mathrm{t \times yr}$ (pessimistic, reduced $^8$B uncertainty) & $\alpha_2$ & $1.75\times10^{-12}$ \\
\hline
\multicolumn{3}{c}{\textbf{XLZD, \eves}} \\
\hline
200~$\mathrm{t \times yr}$  & $\alpha_1$ & $7.94\times10^{-15}$ \\
200~$\mathrm{t \times yr}$  & $\alpha_2$ & $2.95\times10^{-14}$ \\
\hline\hline
\end{tabular}
\caption{The 90\% CL upper limits and projected sensitivities on the invisible neutrino decay parameters obtained in this work.}
\label{tab:inv_decay_limits}
\end{table}

In this paper, we have derived the first constraints on the invisible decay of solar neutrinos using the recently available \cevns~data from current dark matter direct detection experiments. We have also revisited the projected sensitivity of the future XLZD observatory, comparing the reach of coherent elastic neutrino-nucleus scattering (\cevns) and electron-neutrino elastic scattering (\eves).

Focusing on recent \cevns~data collected at the Xe-based facilities XENONnT, PandaX-4T, and LZ, we have shown that these new \cevns~data from solar neutrinos provide constraints on the relevant decay parameter $\alpha_2 = m_2/\tau_2$ that are complementary to, and already comparable in strength to those obtained by SNO~\cite{SNO:2018pvg}.
We have further assessed the projected sensitivity to the same decay parameter through future \cevns~measurements at the proposed multi-ton XLZD experiment. Considering different assumptions on the experimental setup and systematic uncertainties, we find that, unless the theoretical uncertainty on the $^8$B flux is reduced, the sensitivity will not improve substantially with exposure alone.
We have quantified this effect explicitly: an improvement of this uncertainty would yield a corresponding improvement in the sensitivity to $\alpha_2$ of roughly a factor of 2. Therefore, the dominant limitation is no longer detector statistics but the theoretical uncertainty on the solar flux.
Building upon previous studies, we have confirmed that the most powerful channel to probe invisible neutrino decay is, in fact, \eves, which makes the dark matter direct detection experiments sensitive to $pp$ neutrinos, a component of the solar flux composed of both $\nu_1$ and $\nu_2$. The lower energies, a careful treatment of backgrounds, and large event numbers expected at XLZD will allow this channel to probe values of $\alpha_1$ and $\alpha_2$ that are one to two orders of magnitude smaller than current constraints from dedicated solar-neutrino experiments.
We summarize the bounds and sensitivities obtained in this paper in Tab.~\ref{tab:inv_decay_limits}. It should be noted that even though the sensitivity using \cevns~is weaker, it is still a  useful and complementary probe. For example, should a deficit in the $pp$-neutrino flux be observed through \eves, an independent \cevns~measurement would help determine whether the suppression originates from the decay of $\nu_2$ or instead points to $\nu_1$ decay (or a combination of both).

We conclude by noting that the constraints presented in this work are complementary to those obtained from cosmological observations, see for example Refs.~\cite{Escudero:2019gfk,Barenboim:2020vrr,FrancoAbellan:2021hdb}, and the references therein. 
While cosmology is sensitive to the impact of neutrino decay on the thermal history and evolution of large-scale structures, the corresponding limits depend on assumptions regarding the decay products, their interactions, and the epoch at which the decay occurs. 
Solar-neutrino observations, in contrast, directly probe the survival probability of propagating neutrino mass eigenstates over the Sun-Earth baseline and are therefore largely insensitive to the detailed properties of the invisible daughters. 

The recent observation of \cevns~from solar neutrinos has therefore opened a new avenue for precision tests of neutrino properties, with significant discovery potential as the next generation of dark matter detectors comes online.

\section*{Acknowledgments}
C.A.T. is very thankful for the hospitality at IFIC in Valencia, where this work has been finalized. V.D.R. acknowledges financial support by the grant CIDEXG/2022/20 (from Generalitat Valenciana) and by the Spanish grants CNS2023-144124 (MCIN/AEI/10.13039/501100011033 and “Next Generation EU”/PRTR), PID2023-147306NB-I00, and CEX2023-001292-S (MCIU/AEI/10.13039/501100011033). D.K.P. acknowledges funding from the European Union’s Horizon Europe research and innovation programme under the Marie Skłodowska‑Curie Actions grant agreement No.~101198541 (neutrinoSPHERE).

\appendix

\section{Comparison of $pp$ and $^8$B probabilities}
\label{sec:appA}

\begin{figure}[t!]
    \centering
    \includegraphics[width=0.49 \textwidth]{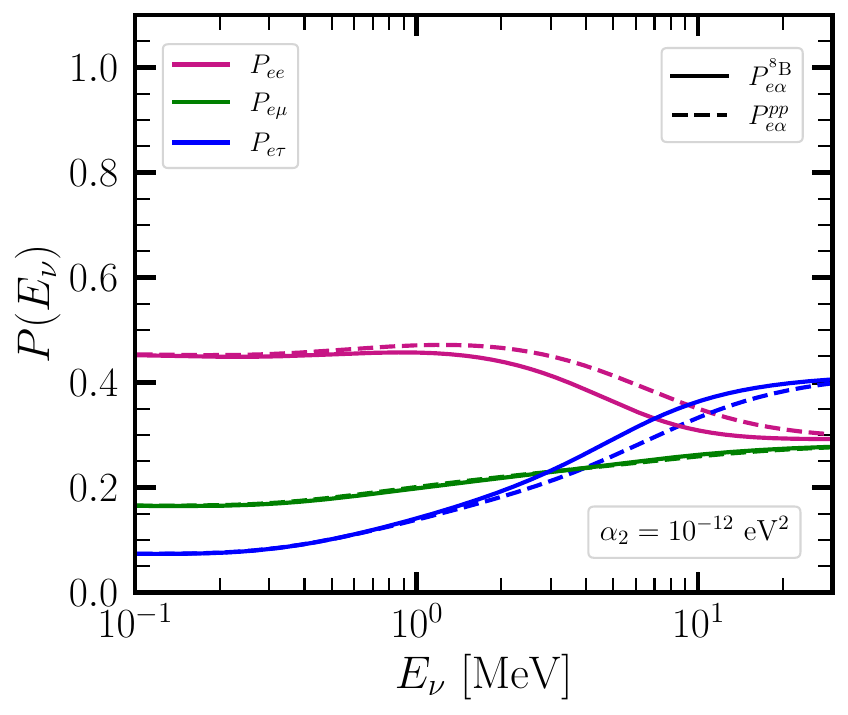} 
    \caption{Comparison of the solar neutrino conversion probabilities in the presence of $\nu_2$ decay, obtained by averaging over the production region using the $\rho_{^{8}\mathrm{B}}(r)$ (solid) and $\rho_{pp}(r)$ (dashed) production probability densities, for a representative value of $\alpha_2 = 10^{-12}$ eV$^2$.}
    \label{fig:pp_vs_8B}
\end{figure}

In this Appendix, we provide further details on the conversion probabilities of solar neutrinos, complementing the discussion in Sec.~\ref{sec:decay}. Specifically, in Fig.~\ref{fig:pp_vs_8B} we compare the probabilities for $pp$ and $^8$B neutrinos, the latter being the same as those shown in Fig.~\ref{fig:probs}. As can be seen, the probabilities associated with these two neutrino sources are very similar. The main difference originates from the fact that the various solar neutrino sources have different radial production probability distributions $\rho_X(r)$ within the interior of the Sun~\cite{Bahcall:2005va}, see, e.g., Eq.~\eqref{eq:rho_integral_prob}. Note that, due to our choice of maximal mixing ($\sin^2\theta_{23}=0.5$) $|U_{\mu1}|^2$ and $|U_{\mu2}|^2$ are quite similar, which leads to an accidental cancellation in $P_{e\mu}$ and therefore the lines for $pp$ neutrinos and ${}^8$B neutrinos are very similar.

\section{Sensitivity of LZ data}
\label{sec:appB}

\begin{figure}[t!]
    \centering
    \includegraphics[width=0.49 \textwidth]{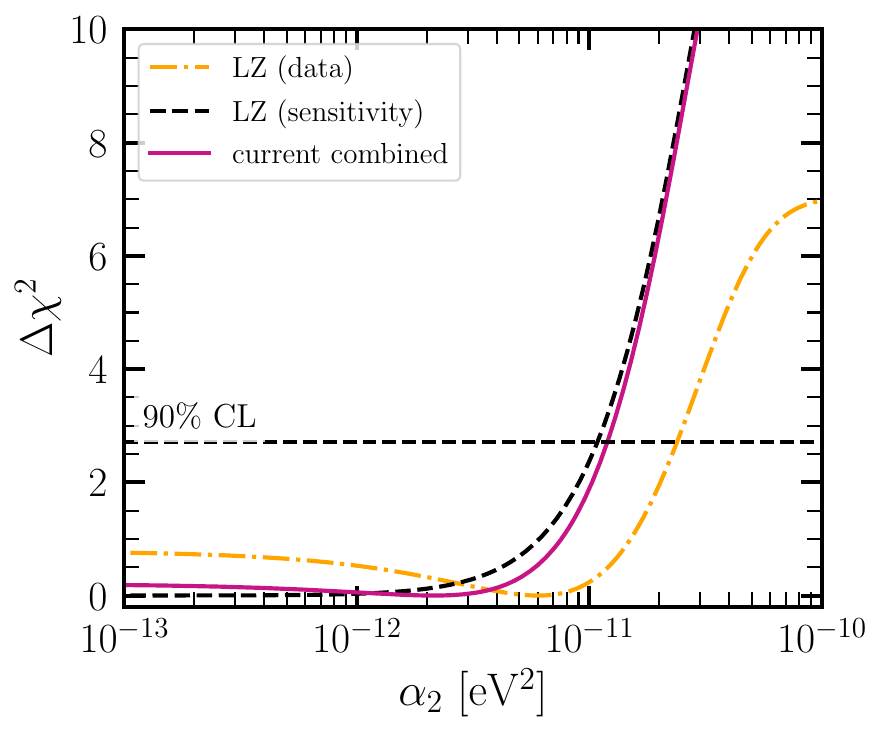}
    \caption{$\chi^2$ profiles of $\alpha_2$  from the analysis of \cevns~data at dark matter direct detection experiments. Sensitivity of LZ (black dashed) in comparison to the constraint coming from the analysis of LZ data (orange dot-dashed) and the current constraint from the combined analysis of XENONnT+PandaX-4T+LZ \cevns~data (magenta solid).}
    \label{fig:LZ_sensitivity}
\end{figure}

As discussed in Sec.~\ref{sec:res_current}, the XENONnT bound of $\alpha_2$ obtained from the analysis of \cevns~data is stronger than that of LZ, even though the latter experiment has observed \cevns~with higher significance. While this may appear counterintuitive, it can be readily understood as follows. The LZ experiment detected fewer \cevns~events compared to the  theoretical expectation, and consequently the best-fit value of the invisible neutrino decay parameter $\alpha_2$ differs from zero (see left panel of Fig.~\ref{fig:CEvNS_bounds}). Indeed, a nonzero value of $\alpha_2$ always leads to a depletion of the expected \cevns~event rate. 
In Fig.~\ref{fig:LZ_sensitivity}, the black dashed line illustrates the $\Delta\chi^2$ profile for $\alpha_2$ assuming that the LZ experiment had observed exactly the theoretically expected number of \cevns~events, referred to as LZ (sensitivity). For comparison, we also reproduce from Fig.~\ref{fig:CEvNS_bounds} (left panel) the result obtained from the analysis of the LZ \cevns~data as well as the result from the combined analysis of the XENONnT, PandaX-4T, and LZ \cevns~data. As expected, the LZ (sensitivity) result is stronger than the current bound which is mainly driven by the XENONnT data.

\bibliography{bibliography}  

@article{Freedman:1973yd,
    author = "Freedman, Daniel Z.",
    title = "{Coherent Neutrino Nucleus Scattering as a Probe of the Weak Neutral Current}",
    reportNumber = "NAL-PUB-73-76-THY, FERMILAB-PUB-73-076-T",
    doi = "10.1103/PhysRevD.9.1389",
    journal = "Phys. Rev. D",
    volume = "9",
    pages = "1389--1392",
    year = "1974"
}

@article{DeRomeri:2025nkx,
    author = "De Romeri, Valentina and Majumdar, Anirban and Papoulias, Dimitrios K. and Srivastava, Rahul",
    title = "{New light mediators and the neutrino fog: Implications from XENONnT nuclear recoil data}",
    eprint = "2512.08853",
    archivePrefix = "arXiv",
    primaryClass = "hep-ph",
    doi = "10.1088/1475-7516/2026/05/093",
    journal = "JCAP",
    volume = "05",
    pages = "093",
    year = "2026"
}

@article{DeRomeri:2024dbv,
    author = "De Romeri, Valentina and Papoulias, Dimitrios K. and Ternes, Christoph A.",
    title = "{Light vector mediators at direct detection experiments}",
    eprint = "2402.05506",
    archivePrefix = "arXiv",
    primaryClass = "hep-ph",
    doi = "10.1007/JHEP05(2024)165",
    journal = "JHEP",
    volume = "05",
    pages = "165",
    year = "2024"
}

@article{AristizabalSierra:2024nwf,
    author = "Aristizabal Sierra, D. and Mishra, N. and Strigari, L.",
    title = "{Implications of first neutrino-induced nuclear recoil measurements in direct detection experiments: Probing nonstandard interaction via CE{\ensuremath{\nu}}NS}",
    eprint = "2409.02003",
    archivePrefix = "arXiv",
    primaryClass = "hep-ph",
    doi = "10.1103/PhysRevD.111.055007",
    journal = "Phys. Rev. D",
    volume = "111",
    number = "5",
    pages = "055007",
    year = "2025"
}

@article{AristizabalSierra:2020zod,
    author = "Aristizabal Sierra, D. and Branada, R. and Miranda, O. G. and Sanchez Garcia, G.",
    title = "{Sensitivity of direct detection experiments to neutrino magnetic dipole moments}",
    eprint = "2008.05080",
    archivePrefix = "arXiv",
    primaryClass = "hep-ph",
    doi = "10.1007/JHEP12(2020)178",
    journal = "JHEP",
    volume = "12",
    pages = "178",
    year = "2020"
}

@article{PandaX:2024muv,
    author = "Bo, Zihao and others",
    collaboration = "PandaX",
    title = "{First Indication of Solar B8 Neutrinos through Coherent Elastic Neutrino-Nucleus Scattering in PandaX-4T}",
    eprint = "2407.10892",
    archivePrefix = "arXiv",
    primaryClass = "hep-ex",
    doi = "10.1103/PhysRevLett.133.191001",
    journal = "Phys. Rev. Lett.",
    volume = "133",
    number = "19",
    pages = "191001",
    year = "2024"
}

@article{XENON:2024ijk,
    author = "Aprile, Elena and others",
    collaboration = "XENON",
    title = "{First Indication of Solar $^8B$~Neutrinos via Coherent Elastic Neutrino-Nucleus Scattering with XENONnT}",
    eprint = "2408.02877",
    archivePrefix = "arXiv",
    primaryClass = "nucl-ex",
    doi = "10.1103/PhysRevLett.133.191002",
    journal = "Phys. Rev. Lett.",
    volume = "133",
    pages = "191002",
    year = "2024"
}

@misc{XENONnT_cevns_data_release,
  author       = {{XENONnT Collaboration}},
  title        = {{CEvNS Data Release}},
  howpublished = {\url{https://github.com/XENONnT/cevns_data_release}}
}

@misc{XENONnT_github,
  author       = {{XENONnT Collaboration}},
  title        = {{A high-Performance Program simuLatEs and fiTs REsponse of xEnon}},
  howpublished = {\url{hhttps://github.com/XENONnT/appletree}}
  }

@article{Amaral:2023tbs,
    author = "Amaral, Dorian W. P. and Cerdeno, David and Cheek, Andrew and Foldenauer, Patrick",
    title = "{A direct detection view of the neutrino NSI landscape}",
    eprint = "2302.12846",
    archivePrefix = "arXiv",
    primaryClass = "hep-ph",
    reportNumber = "IPPP/23/08; IFT-UAM/CSIC-23-19; FT-UAM-23-1",
    doi = "10.1007/JHEP07(2023)071",
    journal = "JHEP",
    volume = "07",
    pages = "071",
    year = "2023"
}

@article{Aalbers:2022dzr,
    author = "Aalbers, J. and others",
    title = "{A next-generation liquid xenon observatory for dark matter and neutrino physics}",
    eprint = "2203.02309",
    archivePrefix = "arXiv",
    primaryClass = "physics.ins-det",
    reportNumber = "INT-PUB-22-003, FERMILAB-PUB-22-112-PPD-QIS-T",
    doi = "10.1088/1361-6471/ac841a",
    journal = "J. Phys. G",
    volume = "50",
    number = "1",
    pages = "013001",
    year = "2023"
}

@article{Li:2024iij,
    author = "Li, Gang and Song, Chuan-Qiang and Tang, Feng-Jie and Yu, Jiang-Hao",
    title = "{Constraints on neutrino nonstandard interactions from COHERENT, PandaX-4T and XENONnT}",
    eprint = "2409.04703",
    archivePrefix = "arXiv",
    primaryClass = "hep-ph",
    doi = "10.1103/PhysRevD.111.035002",
    journal = "Phys. Rev. D",
    volume = "111",
    number = "3",
    pages = "035002",
    year = "2025"
}

@article{Xia:2024ytb,
    author = "Xia, Shuo-yu",
    title = "{Measuring solar neutrino fluxes in direct detection experiments in the presence of light mediators}",
    eprint = "2410.01167",
    archivePrefix = "arXiv",
    primaryClass = "hep-ph",
    doi = "10.1016/j.nuclphysb.2024.116738",
    journal = "Nucl. Phys. B",
    volume = "1009",
    pages = "116738",
    year = "2024"
}

@article{deSalas:2020pgw,
    author = "de Salas, P. F. and Forero, D. V. and Gariazzo, S. and Mart\'\i{}nez-Mirav\'e, P. and Mena, O. and Ternes, C. A. and T\'ortola, M. and Valle, J. W. F.",
    title = "{2020 global reassessment of the neutrino oscillation picture}",
    eprint = "2006.11237",
    archivePrefix = "arXiv",
    primaryClass = "hep-ph",
    doi = "10.1007/JHEP02(2021)071",
    journal = "JHEP",
    volume = "02",
    pages = "071",
    year = "2021"
}

@article{XENON:2022ltv,
    author = "Aprile, E. and others",
    collaboration = "XENON",
    title = "{Search for New Physics in Electronic Recoil Data from XENONnT}",
    eprint = "2207.11330",
    archivePrefix = "arXiv",
    primaryClass = "hep-ex",
    doi = "10.1103/PhysRevLett.129.161805",
    journal = "Phys. Rev. Lett.",
    volume = "129",
    number = "16",
    pages = "161805",
    year = "2022"
}

@article{Giunti:2023yha,
    author = "Giunti, Carlo and Ternes, Christoph A.",
    title = "{Testing neutrino electromagnetic properties at current and future dark matter experiments}",
    eprint = "2309.17380",
    archivePrefix = "arXiv",
    primaryClass = "hep-ph",
    doi = "10.1103/PhysRevD.108.095044",
    journal = "Phys. Rev. D",
    volume = "108",
    number = "9",
    pages = "095044",
    year = "2023"
}

@article{ParticleDataGroup:2024cfk,
    author = "Navas, S. and others",
    collaboration = "Particle Data Group",
    title = "{Review of particle physics}",
    doi = "10.1103/PhysRevD.110.030001",
    journal = "Phys. Rev. D",
    volume = "110",
    number = "3",
    pages = "030001",
    year = "2024"
}

@article{Klein:1999qj,
    author = "Klein, Spencer and Nystrand, Joakim",
    title = "{Exclusive vector meson production in relativistic heavy ion collisions}",
    eprint = "hep-ph/9902259",
    archivePrefix = "arXiv",
    reportNumber = "LBNL-42768, LBL-42768",
    doi = "10.1103/PhysRevC.60.014903",
    journal = "Phys. Rev. C",
    volume = "60",
    pages = "014903",
    year = "1999"
}

@article{Alonso-Gonzalez:2023tgm,
    author = "Alonso-Gonz\'alez, D. and Amaral, D. W. P. and Bariego-Quintana, A. and Cerde\~no, D. and de los Rios, M.",
    title = "{Measuring the sterile neutrino mass in spallation source and direct detection experiments}",
    eprint = "2307.05176",
    archivePrefix = "arXiv",
    primaryClass = "hep-ph",
    reportNumber = "IFT-UAM/CSIC-23-89",
    doi = "10.1007/JHEP12(2023)096",
    journal = "JHEP",
    volume = "12",
    pages = "096",
    year = "2023"
}

@article{Abdullah:2022zue,
    author = "Abdullah, M. and others",
    title = "{Coherent elastic neutrino-nucleus scattering: Terrestrial and astrophysical applications}",
    eprint = "2203.07361",
    archivePrefix = "arXiv",
    primaryClass = "hep-ph",
    month = "3",
    year = "2022"
}

@article{Barranco:2005yy,
    author = "Barranco, J. and Miranda, O. G. and Rashba, T. I.",
    title = "{Probing new physics with coherent neutrino scattering off nuclei}",
    eprint = "hep-ph/0508299",
    archivePrefix = "arXiv",
    reportNumber = "MPP-2005-85",
    doi = "10.1088/1126-6708/2005/12/021",
    journal = "JHEP",
    volume = "12",
    pages = "021",
    year = "2005"
}

@article{Cerdeno:2016sfi,
    author = "Cerde\~no, David G. and Fairbairn, Malcolm and Jubb, Thomas and Machado, Pedro A. N. and Vincent, Aaron C. and B\oe{}hm, C\'eline",
    title = "{Physics from solar neutrinos in dark matter direct detection experiments}",
    eprint = "1604.01025",
    archivePrefix = "arXiv",
    primaryClass = "hep-ph",
    reportNumber = "IFT-UAM-CSIC-16-031, FTUAM-16-12, IPPP-16-27, DCTP-16-54, KCL-PH-TH-2016-19",
    doi = "10.1007/JHEP09(2016)048",
    journal = "JHEP",
    volume = "05",
    pages = "118",
    year = "2016",
    note = "[Erratum: JHEP 09, 048 (2016)]"
}

@article{Dutta:2017nht,
    author = "Dutta, Bhaskar and Liao, Shu and Strigari, Louis E. and Walker, Joel W.",
    title = "{Non-standard interactions of solar neutrinos in dark matter experiments}",
    eprint = "1705.00661",
    archivePrefix = "arXiv",
    primaryClass = "hep-ph",
    reportNumber = "MI-TH-1724",
    doi = "10.1016/j.physletb.2017.08.031",
    journal = "Phys. Lett. B",
    volume = "773",
    pages = "242--246",
    year = "2017"
}

@article{Gelmini:2018gqa,
    author = "Gelmini, Graciela B. and Takhistov, Volodymyr and Witte, Samuel J.",
    title = "{Geoneutrinos in Large Direct Detection Experiments}",
    eprint = "1812.05550",
    archivePrefix = "arXiv",
    primaryClass = "hep-ph",
    doi = "10.1103/PhysRevD.99.093009",
    journal = "Phys. Rev. D",
    volume = "99",
    number = "9",
    pages = "093009",
    year = "2019"
}

@article{Essig:2018tss,
    author = "Essig, Rouven and Sholapurkar, Mukul and Yu, Tien-Tien",
    title = "{Solar Neutrinos as a Signal and Background in Direct-Detection  Experiments Searching for Sub-GeV Dark Matter With Electron Recoils}",
    eprint = "1801.10159",
    archivePrefix = "arXiv",
    primaryClass = "hep-ph",
    reportNumber = "YITB-SB-17-36, YITP-SB-17-36, CERN-TH-2017-194",
    doi = "10.1103/PhysRevD.97.095029",
    journal = "Phys. Rev. D",
    volume = "97",
    number = "9",
    pages = "095029",
    year = "2018"
}

@article{Amaral:2020tga,
    author = "Amaral, Dorian Warren Praia do and Cerdeno, David G. and Foldenauer, Patrick and Reid, Elliott",
    title = "{Solar neutrino probes of the muon anomalous magnetic moment in the gauged $ \mathrm{U}{(1)}_{L_{\mu }-{L}_{\tau }} $}",
    eprint = "2006.11225",
    archivePrefix = "arXiv",
    primaryClass = "hep-ph",
    reportNumber = "IPPP/20/24, IFT-UAM/CSIC-20-70",
    doi = "10.1007/JHEP12(2020)155",
    journal = "JHEP",
    volume = "12",
    pages = "155",
    year = "2020"
}

@article{Dutta:2020che,
    author = "Dutta, Bhaskar and Lang, Rafael F. and Liao, Shu and Sinha, Samiran and Strigari, Louis and Thompson, Adrian",
    title = "{A global analysis strategy to resolve neutrino NSI degeneracies with scattering and oscillation data}",
    eprint = "2002.03066",
    archivePrefix = "arXiv",
    primaryClass = "hep-ph",
    reportNumber = "MI-TH-204",
    doi = "10.1007/JHEP09(2020)106",
    journal = "JHEP",
    volume = "09",
    pages = "106",
    year = "2020"
}

@article{Amaral:2021rzw,
    author = "Amaral, D. W. P. and Cerdeno, D. G. and Cheek, A. and Foldenauer, P.",
    title = "{Confirming $U(1)_{L_\mu -L_{\tau }}$ as a solution for $(g-2)_\mu $ with neutrinos}",
    eprint = "2104.03297",
    archivePrefix = "arXiv",
    primaryClass = "hep-ph",
    reportNumber = "IPPP/20/94, IFT-UAM/CSIC-20-35, CP3-21-11",
    doi = "10.1140/epjc/s10052-021-09670-z",
    journal = "Eur. Phys. J. C",
    volume = "81",
    number = "10",
    pages = "861",
    year = "2021"
}

@article{deGouvea:2021ymm,
    author = "de Gouv\^ea, Andr\'e and McGinness, Emma and Martinez-Soler, Ivan and Perez-Gonzalez, Yuber F.",
    title = "{pp solar neutrinos at DARWIN}",
    eprint = "2111.02421",
    archivePrefix = "arXiv",
    primaryClass = "hep-ph",
    reportNumber = "NUHEP-TH/21-17, FERMILAB-PUB-21-560-T, IPPP/21/46",
    doi = "10.1103/PhysRevD.106.096017",
    journal = "Phys. Rev. D",
    volume = "106",
    number = "9",
    pages = "096017",
    year = "2022"
}

@article{Monroe:2007xp,
    author = "Monroe, Jocelyn and Fisher, Peter",
    title = "{Neutrino Backgrounds to Dark Matter Searches}",
    eprint = "0706.3019",
    archivePrefix = "arXiv",
    primaryClass = "astro-ph",
    doi = "10.1103/PhysRevD.76.033007",
    journal = "Phys. Rev. D",
    volume = "76",
    pages = "033007",
    year = "2007"
}

@article{Vergados:2008jp,
    author = "Vergados, J. D. and Ejiri, H.",
    title = "{Can Solar Neutrinos be a Serious Background in Direct Dark Matter Searches?}",
    eprint = "0805.2583",
    archivePrefix = "arXiv",
    primaryClass = "hep-ph",
    doi = "10.1016/j.nuclphysb.2008.06.004",
    journal = "Nucl. Phys. B",
    volume = "804",
    pages = "144--159",
    year = "2008"
}

@article{Strigari:2009bq,
    author = "Strigari, Louis E.",
    title = "{Neutrino Coherent Scattering Rates at Direct Dark Matter Detectors}",
    eprint = "0903.3630",
    archivePrefix = "arXiv",
    primaryClass = "astro-ph.CO",
    doi = "10.1088/1367-2630/11/10/105011",
    journal = "New J. Phys.",
    volume = "11",
    pages = "105011",
    year = "2009"
}

@article{Bahcall:2005va,
    author = "Bahcall, John N. and Serenelli, Aldo M. and Basu, Sarbani",
    title = "{10,000 standard solar models: a Monte Carlo simulation}",
    eprint = "astro-ph/0511337",
    archivePrefix = "arXiv",
    doi = "10.1086/504043",
    journal = "Astrophys. J. Suppl.",
    volume = "165",
    pages = "400--431",
    year = "2006"
}

@article{Billard:2013qya,
    author = "Billard, J. and Strigari, L. and Figueroa-Feliciano, E.",
    title = "{Implication of neutrino backgrounds on the reach of next generation dark matter direct detection experiments}",
    eprint = "1307.5458",
    archivePrefix = "arXiv",
    primaryClass = "hep-ph",
    doi = "10.1103/PhysRevD.89.023524",
    journal = "Phys. Rev. D",
    volume = "89",
    number = "2",
    pages = "023524",
    year = "2014"
}

@article{OHare:2021utq,
    author = "O'Hare, Ciaran A. J.",
    title = "{New Definition of the Neutrino Floor for Direct Dark Matter Searches}",
    eprint = "2109.03116",
    archivePrefix = "arXiv",
    primaryClass = "hep-ph",
    doi = "10.1103/PhysRevLett.127.251802",
    journal = "Phys. Rev. Lett.",
    volume = "127",
    number = "25",
    pages = "251802",
    year = "2021"
}

@article{Chen:2016eab,
    author = "Chen, Jiunn-Wei and Chi, Hsin-Chang and Liu, C. -P. and Wu, Chih-Pan",
    title = "{Low-energy electronic recoil in xenon detectors by solar neutrinos}",
    eprint = "1610.04177",
    archivePrefix = "arXiv",
    primaryClass = "hep-ex",
    reportNumber = "NCTS-ECP-1601, MIT-CTP-4843",
    doi = "10.1016/j.physletb.2017.10.029",
    journal = "Phys. Lett. B",
    volume = "774",
    pages = "656--661",
    year = "2017"
}

@article{Cadeddu:2020lky,
    author = "Cadeddu, M. and Dordei, F. and Giunti, C. and Li, Y. F. and Picciau, E. and Zhang, Y. Y.",
    title = "{Physics results from the first COHERENT observation of coherent elastic neutrino-nucleus scattering in argon and their combination with cesium-iodide data}",
    eprint = "2005.01645",
    archivePrefix = "arXiv",
    primaryClass = "hep-ph",
    doi = "10.1103/PhysRevD.102.015030",
    journal = "Phys. Rev. D",
    volume = "102",
    number = "1",
    pages = "015030",
    year = "2020"
}

@article{Harnik:2012ni,
    author = "Harnik, Roni and Kopp, Joachim and Machado, Pedro A. N.",
    title = "{Exploring nu Signals in Dark Matter Detectors}",
    eprint = "1202.6073",
    archivePrefix = "arXiv",
    primaryClass = "hep-ph",
    reportNumber = "FERMILAB-PUB-12-048-T",
    doi = "10.1088/1475-7516/2012/07/026",
    journal = "JCAP",
    volume = "07",
    pages = "026",
    year = "2012"
}

@article{Suliga:2020jfa,
    author = "Suliga, Anna M. and Tamborra, Irene",
    title = "{Astrophysical constraints on nonstandard coherent neutrino-nucleus scattering}",
    eprint = "2010.14545",
    archivePrefix = "arXiv",
    primaryClass = "hep-ph",
    doi = "10.1103/PhysRevD.103.083002",
    journal = "Phys. Rev. D",
    volume = "103",
    number = "8",
    pages = "083002",
    year = "2021"
}

@article{DeRomeri:2024hvc,
    author = "De Romeri, Valentina and Papoulias, Dimitrios K. and Sanchez Garcia, Gonzalo and Ternes, Christoph A. and T{\'o}rtola, Mariam",
    title = "{Neutrino electromagnetic properties and sterile dipole portal in light of the first solar CE{\ensuremath{\nu}}NS~data}",
    eprint = "2412.14991",
    archivePrefix = "arXiv",
    primaryClass = "hep-ph",
    doi = "10.1088/1475-7516/2025/05/080",
    journal = "JCAP",
    volume = "05",
    pages = "080",
    year = "2025"
}

@article{Gehrlein:2025isp,
    author = "Gehrlein, Julia and Kushwaha, Tanmay",
    title = "{Testing the dark side of neutrino oscillations with the solar neutrino fog at dark matter experiments}",
    eprint = "2508.14166",
    archivePrefix = "arXiv",
    primaryClass = "hep-ph",
    doi = "10.1103/g74v-lr8t",
    journal = "Phys. Rev. D",
    volume = "112",
    number = "9",
    pages = "095032",
    year = "2025"
}

@article{AtzoriCorona:2025gyz,
    author = "Atzori Corona, M. and Cadeddu, M. and Cargioli, N. and Dordei, F. and Sestu, M.",
    title = "{When backgrounds become signals: neutrino interactions in xenon-based dark matter detectors}",
    eprint = "2509.22178",
    archivePrefix = "arXiv",
    primaryClass = "hep-ph",
    doi = "10.1088/1475-7516/2026/05/065",
    journal = "JCAP",
    volume = "05",
    pages = "065",
    year = "2026"
}

@article{Blanco-Mas:2024ale,
    author = "Blanco-Mas, Pablo and Coloma, Pilar and Herrera, Gonzalo and Huber, Patrick and Kopp, Joachim and Shoemaker, Ian M. and Tabrizi, Zahra",
    title = "{Clarity through the neutrino fog: constraining new forces in dark matter detectors}",
    eprint = "2411.14206",
    archivePrefix = "arXiv",
    primaryClass = "hep-ph",
    reportNumber = "IFT-UAM/CSIC-24-164",
    doi = "10.1007/JHEP08(2025)043",
    journal = "JHEP",
    volume = "08",
    pages = "043",
    year = "2025"
}

@article{Capozzi:2025wyn,
    author = "Capozzi, Francesco and Giar{\`e}, William and Lisi, Eligio and Marrone, Antonio and Melchiorri, Alessandro and Palazzo, Antonio",
    title = "{Neutrino masses and mixing: Entering the era of subpercent precision}",
    eprint = "2503.07752",
    archivePrefix = "arXiv",
    primaryClass = "hep-ph",
    doi = "10.1103/PhysRevD.111.093006",
    journal = "Phys. Rev. D",
    volume = "111",
    number = "9",
    pages = "093006",
    year = "2025"
}

@article{Esteban:2024eli,
    author = "Esteban, Ivan and Gonzalez-Garcia, M. C. and Maltoni, Michele and Martinez-Soler, Ivan and Pinheiro, Jo{\~a}o Paulo and Schwetz, Thomas",
    title = "{NuFit-6.0: updated global analysis of three-flavor neutrino oscillations}",
    eprint = "2410.05380",
    archivePrefix = "arXiv",
    primaryClass = "hep-ph",
    reportNumber = "IFT-UAM/CSIC-24-140, YITP-SB-2024-24, IPPP/24/64, IPPP/24/64, IFT-UAM/CSIC-24-140, YITP-SB-2024-24",
    doi = "10.1007/JHEP12(2024)216",
    journal = "JHEP",
    volume = "12",
    pages = "216",
    year = "2024"
}

@article{JUNO:2025gmd,
    author = "Abusleme, Angel and others",
    collaboration = "JUNO",
    title = "{Measurement of reactor neutrino oscillation with the first JUNO data}",
    eprint = "2511.14593",
    archivePrefix = "arXiv",
    primaryClass = "hep-ex",
    doi = "10.1038/s41586-026-10538-z",
    journal = "Nature",
    volume = "654",
    number = "8118",
    pages = "343--348",
    year = "2026"
}

@article{LZ:2025igz,
    author = "Akerib, D. S. and others",
    collaboration = "LZ",
    title = "{Searches for Light Dark Matter and Evidence of Coherent Elastic Neutrino-Nucleus Scattering of Solar Neutrinos with the LUX-ZEPLIN (LZ) Experiment}",
    eprint = "2512.08065",
    archivePrefix = "arXiv",
    primaryClass = "hep-ex",
    month = "12",
    year = "2025"
}

@article{DeRomeri:2024iaw,
    author = "De Romeri, Valentina and Papoulias, Dimitrios K. and Ternes, Christoph A.",
    title = "{Bounds on new neutrino interactions from the first CE{\ensuremath{\nu}}NS data at direct detection experiments}",
    eprint = "2411.11749",
    archivePrefix = "arXiv",
    primaryClass = "hep-ph",
    doi = "10.1088/1475-7516/2025/05/012",
    journal = "JCAP",
    volume = "05",
    pages = "012",
    year = "2025"
}

@article{AtzoriCorona:2025xwr,
    author = "Atzori Corona, Mattia and Cadeddu, Matteo and Cargioli, Nicola and Dordei, Francesca and Giunti, Carlo and Ternes, Christoph A.",
    title = "{Standard Model Tested with Neutrinos}",
    eprint = "2504.05272",
    archivePrefix = "arXiv",
    primaryClass = "hep-ph",
    doi = "10.1103/dplq-dvc8",
    journal = "Phys. Rev. Lett.",
    volume = "135",
    number = "23",
    pages = "231803",
    year = "2025"
}

@article{XENON:2026ydt,
    author = "Aprile, E. and others",
    collaboration = "XENON",
    title = "{Probing the Solar $^8$B Neutrino Fog with XENONnT}",
    eprint = "2604.06002",
    archivePrefix = "arXiv",
    primaryClass = "hep-ex",
    month = "4",
    year = "2026"
}

@article{DeRomeri:2026prc,
    author = "De Romeri, Valentina and Papoulias, Dimitrios K. and Pompa, Federica and Sanchez Garcia, Gonzalo and Ternes, Christoph A.",
    title = "{Testing light and heavy vector mediators with solar CE$ν$NS measurements}",
    eprint = "2603.00554",
    archivePrefix = "arXiv",
    primaryClass = "hep-ph",
    month = "2",
    year = "2026"
}

@article{Pompa:2023yzg,
    author = "Pompa, Federica and Mena, Olga",
    title = "{How long do neutrinos live and how much do they weigh?}",
    eprint = "2310.05474",
    archivePrefix = "arXiv",
    primaryClass = "hep-ph",
    doi = "10.1140/epjc/s10052-024-12499-x",
    journal = "Eur. Phys. J. C",
    volume = "84",
    number = "2",
    pages = "134",
    year = "2024"
}

@article{Kelly:2026avh,
    author = "Kelly, Kevin J. and Mishra, Nityasa and Strigari, Louis E.",
    title = "{Sterile Neutrino Mixing Parameters from Solar-Neutrino Coherent Scattering}",
    eprint = "2605.22935",
    archivePrefix = "arXiv",
    primaryClass = "hep-ph",
    reportNumber = "MI-HET-884",
    month = "5",
    year = "2026"
}

@article{Frieman:1987as,
    author = "Frieman, Joshua A. and Haber, Howard E. and Freese, Katherine",
    title = "{Neutrino Mixing, Decays and Supernova Sn1987a}",
    reportNumber = "SLAC-PUB-4261, SCIPP-87-90, NSF-ITP-87-53",
    doi = "10.1016/0370-2693(88)91120-3",
    journal = "Phys. Lett. B",
    volume = "200",
    pages = "115--121",
    year = "1988"
}

@article{Ivanez-Ballesteros:2023lqa,
    author = "Iv{\'a}{\~n}ez-Ballesteros, Pilar and Volpe, Maria Cristina",
    title = "{SN1987A and neutrino non-radiative decay}",
    eprint = "2307.03549",
    archivePrefix = "arXiv",
    primaryClass = "hep-ph",
    doi = "10.1016/j.physletb.2023.138252",
    journal = "Phys. Lett. B",
    volume = "847",
    pages = "138252",
    year = "2023"
}

@article{Ternes:2024qui,
    author = "Ternes, Christoph A. and Pagliaroli, Giulia",
    title = "{Invisible neutrino decay at long-baseline neutrino oscillation experiments}",
    eprint = "2401.14316",
    archivePrefix = "arXiv",
    primaryClass = "hep-ph",
    doi = "10.1103/PhysRevD.109.L071701",
    journal = "Phys. Rev. D",
    volume = "109",
    number = "7",
    pages = "L071701",
    year = "2024"
}

@article{Martinez-Mirave:2024hfd,
    author = "Mart{\'\i}nez-Mirav{\'e}, Pablo and Tamborra, Irene and T{\'o}rtola, Mariam",
    title = "{The Sun and core-collapse supernovae are leading probes of the neutrino lifetime}",
    eprint = "2402.00116",
    archivePrefix = "arXiv",
    primaryClass = "astro-ph.HE",
    doi = "10.1088/1475-7516/2024/05/002",
    journal = "JCAP",
    volume = "05",
    pages = "002",
    year = "2024"
}

@article{Gonzalez-Garcia:2008mgl,
    author = "Gonzalez-Garcia, M. C. and Maltoni, M.",
    title = "{Status of Oscillation plus Decay of Atmospheric and Long-Baseline Neutrinos}",
    eprint = "0802.3699",
    archivePrefix = "arXiv",
    primaryClass = "hep-ph",
    reportNumber = "YITP-SB-08-03, IFT-UAM-CSIC-08-10",
    doi = "10.1016/j.physletb.2008.04.041",
    journal = "Phys. Lett. B",
    volume = "663",
    pages = "405--409",
    year = "2008"
}

@article{Berryman:2014qha,
    author = "Berryman, Jeffrey M. and de Gouvea, Andre and Hernandez, Daniel",
    title = "{Solar Neutrinos and the Decaying Neutrino Hypothesis}",
    eprint = "1411.0308",
    archivePrefix = "arXiv",
    primaryClass = "hep-ph",
    reportNumber = "NUHEP-TH-14-08",
    doi = "10.1103/PhysRevD.92.073003",
    journal = "Phys. Rev. D",
    volume = "92",
    number = "7",
    pages = "073003",
    year = "2015"
}

@article{Davis:1968cp,
    author = "Davis, Jr., Raymond and Harmer, Don S. and Hoffman, Kenneth C.",
    title = "{Search for neutrinos from the sun}",
    doi = "10.1103/PhysRevLett.20.1205",
    journal = "Phys. Rev. Lett.",
    volume = "20",
    pages = "1205--1209",
    year = "1968"
}

@article{Kamiokande-II:1989hkh,
    author = "Hirata, K. S. and others",
    collaboration = "Kamiokande-II",
    title = "{Observation of B-8 Solar Neutrinos in the Kamiokande-II Detector}",
    reportNumber = "ICR-188-89-5, KEK-Preprint-89-2, KOBE-AP-89-04, OULNS-89-01, UPR-0165E",
    doi = "10.1103/PhysRevLett.63.16",
    journal = "Phys. Rev. Lett.",
    volume = "63",
    pages = "16",
    year = "1989"
}

@article{GALLEX:1992gcp,
    author = "Anselmann, P. and others",
    collaboration = "GALLEX",
    title = "{Solar neutrinos observed by GALLEX at Gran Sasso.}",
    doi = "10.1016/0370-2693(92)91521-A",
    journal = "Phys. Lett. B",
    volume = "285",
    pages = "376--389",
    year = "1992"
}

@article{Abdurashitov:1996dp,
    author = "Abdurashitov, Dzh. N. and others",
    title = "{The Russian-American gallium experiment (SAGE) Cr neutrino source measurement}",
    doi = "10.1103/PhysRevLett.77.4708",
    journal = "Phys. Rev. Lett.",
    volume = "77",
    pages = "4708--4711",
    year = "1996"
}

@article{Super-Kamiokande:1998zvz,
    author = "Fukuda, Y. and others",
    collaboration = "Super-Kamiokande",
    title = "{Measurement of the solar neutrino energy spectrum using neutrino electron scattering}",
    eprint = "hep-ex/9812011",
    archivePrefix = "arXiv",
    reportNumber = "ICRR-REPORT-442-98-38, NGTHEP-98-02, TIT-HPE-99-02, KEK-PREPRINT-98-219, UCI-99-03, SBHEP-98-9, LSU-HEPA-2-99",
    doi = "10.1103/PhysRevLett.82.2430",
    journal = "Phys. Rev. Lett.",
    volume = "82",
    pages = "2430--2434",
    year = "1999"
}

@article{SNO:2002tuh,
    author = "Ahmad, Q. R. and others",
    collaboration = "SNO",
    title = "{Direct evidence for neutrino flavor transformation from neutral current interactions in the Sudbury Neutrino Observatory}",
    eprint = "nucl-ex/0204008",
    archivePrefix = "arXiv",
    doi = "10.1103/PhysRevLett.89.011301",
    journal = "Phys. Rev. Lett.",
    volume = "89",
    pages = "011301",
    year = "2002"
}

@article{Wolfenstein:1977ue,
    author = "Wolfenstein, L.",
    title = "{Neutrino Oscillations in Matter}",
    reportNumber = "COO-3066-102",
    doi = "10.1103/PhysRevD.17.2369",
    journal = "Phys. Rev. D",
    volume = "17",
    pages = "2369--2374",
    year = "1978"
}

@article{Mikheyev:1985zog,
    author = "Mikheyev, S. P. and Smirnov, A. Yu.",
    title = "{Resonance Amplification of Oscillations in Matter and Spectroscopy of Solar Neutrinos}",
    journal = "Sov. J. Nucl. Phys.",
    volume = "42",
    pages = "913--917",
    year = "1985"
}

@article{Maltoni:2015kca,
    author = "Maltoni, Michele and Smirnov, Alexei Yu.",
    title = "{Solar neutrinos and neutrino physics}",
    eprint = "1507.05287",
    archivePrefix = "arXiv",
    primaryClass = "hep-ph",
    reportNumber = "IFT-UAM-CSIC-15-069",
    doi = "10.1140/epja/i2016-16087-0",
    journal = "Eur. Phys. J. A",
    volume = "52",
    number = "4",
    pages = "87",
    year = "2016"
}

@article{Villante:2020adi,
    author = "Villante, Francesco L. and Serenelli, Aldo",
    title = "{The relevance of nuclear reactions for Standard Solar Models construction}",
    eprint = "2101.03077",
    archivePrefix = "arXiv",
    primaryClass = "astro-ph.SR",
    doi = "10.3389/fspas.2020.618356",
    journal = "Front. Astron. Space Sci.",
    volume = "7",
    pages = "112",
    year = "2021"
}

@article{DARWIN:2020bnc,
    author = "Aalbers, J. and others",
    collaboration = "DARWIN",
    title = "{Solar neutrino detection sensitivity in DARWIN via electron scattering}",
    eprint = "2006.03114",
    archivePrefix = "arXiv",
    primaryClass = "physics.ins-det",
    doi = "10.1140/epjc/s10052-020-08602-7",
    journal = "Eur. Phys. J. C",
    volume = "80",
    number = "12",
    pages = "1133",
    year = "2020"
}

@article{Celestino-Ramirez:2025snn,
    author = "Celestino-Ram{\'\i}rez, Jes{\'u}s Miguel and Escrihuela, F. J. and Flores, L. J. and Miranda, O. G. and S{\'a}nchez-V{\'e}lez, R.",
    title = "{Searching for generalized neutrino interactions in direct detection experiments with E{\ensuremath{\nu}}ES}",
    eprint = "2510.17027",
    archivePrefix = "arXiv",
    primaryClass = "hep-ph",
    doi = "10.1007/JHEP05(2026)071",
    journal = "JHEP",
    volume = "05",
    pages = "071",
    year = "2026"
}

@article{Pagliaroli:2015rca,
    author = "Pagliaroli, G. and Palladino, A. and Villante, F. L. and Vissani, F.",
    title = "{Testing nonradiative neutrino decay scenarios with IceCube data}",
    eprint = "1506.02624",
    archivePrefix = "arXiv",
    primaryClass = "hep-ph",
    doi = "10.1103/PhysRevD.92.113008",
    journal = "Phys. Rev. D",
    volume = "92",
    number = "11",
    pages = "113008",
    year = "2015"
}

@article{Denton:2018aml,
    author = "Denton, Peter B. and Tamborra, Irene",
    title = "{Invisible Neutrino Decay Could Resolve IceCube\textquoteright{}s Track and Cascade Tension}",
    eprint = "1805.05950",
    archivePrefix = "arXiv",
    primaryClass = "hep-ph",
    doi = "10.1103/PhysRevLett.121.121802",
    journal = "Phys. Rev. Lett.",
    volume = "121",
    number = "12",
    pages = "121802",
    year = "2018"
}

@article{Choubey:2018cfz,
    author = "Choubey, Sandhya and Dutta, Debajyoti and Pramanik, Dipyaman",
    title = "{Invisible neutrino decay in the light of NOvA and T2K data}",
    eprint = "1805.01848",
    archivePrefix = "arXiv",
    primaryClass = "hep-ph",
    doi = "10.1007/JHEP08(2018)141",
    journal = "JHEP",
    volume = "08",
    pages = "141",
    year = "2018"
}

@article{Lindner:2001fx,
    author = "Lindner, Manfred and Ohlsson, Tommy and Winter, Walter",
    title = "{A Combined treatment of neutrino decay and neutrino oscillations}",
    eprint = "hep-ph/0103170",
    archivePrefix = "arXiv",
    reportNumber = "TUM-HEP-408-01",
    doi = "10.1016/S0550-3213(01)00237-1",
    journal = "Nucl. Phys. B",
    volume = "607",
    pages = "326--354",
    year = "2001"
}

@article{Abrahao:2015rba,
    author = "Abrah\~ao, Thamys and Minakata, Hisakazu and Nunokawa, Hiroshi and Quiroga, Alexander A.",
    title = "{Constraint on Neutrino Decay with Medium-Baseline Reactor Neutrino Oscillation Experiments}",
    eprint = "1506.02314",
    archivePrefix = "arXiv",
    primaryClass = "hep-ph",
    reportNumber = "INT-PUB-15-024",
    doi = "10.1007/JHEP11(2015)001",
    journal = "JHEP",
    volume = "11",
    pages = "001",
    year = "2015"
}

@article{Ghoshal:2020hyo,
    author = "Ghoshal, Anish and Giarnetti, Alessio and Meloni, Davide",
    title = "{Neutrino Invisible Decay at DUNE: a multi-channel analysis}",
    eprint = "2003.09012",
    archivePrefix = "arXiv",
    primaryClass = "hep-ph",
    doi = "10.1088/1361-6471/abdfab",
    journal = "J. Phys. G",
    volume = "48",
    number = "5",
    pages = "055004",
    year = "2021"
}

@article{Chattopadhyay:2022ftv,
    author = "Chattopadhyay, Dibya S. and Chakraborty, Kaustav and Dighe, Amol and Goswami, Srubabati",
    title = "{Analytic treatment of 3-flavor neutrino oscillation and decay in matter}",
    eprint = "2204.05803",
    archivePrefix = "arXiv",
    primaryClass = "hep-ph",
    reportNumber = "TIFR/TH/22-17",
    doi = "10.1007/JHEP01(2023)051",
    journal = "JHEP",
    volume = "01",
    pages = "051",
    year = "2023"
}

@article{Banerjee:2023sxj,
    author = "Banerjee, Rajrupa and Sharma, Kiran and Patra, Sudhanwa and Panigrahi, Prasanta K.",
    title = "{Geometrical Interpretation of Neutrino Oscillation with decay}",
    journal = "",
    eprint = "2312.08178",
    archivePrefix = "arXiv",
    primaryClass = "hep-ph",
    month = "12",
    year = "2023"
}

@article{Chattopadhyay:2021eba,
    author = "Chattopadhyay, Dibya S. and Chakraborty, Kaustav and Dighe, Amol and Goswami, Srubabati and Lakshmi, S. M.",
    title = "{Neutrino Propagation When Mass Eigenstates and Decay Eigenstates Mismatch}",
    eprint = "2111.13128",
    archivePrefix = "arXiv",
    primaryClass = "hep-ph",
    reportNumber = "TIFR/TH/21-21",
    doi = "10.1103/PhysRevLett.129.011802",
    journal = "Phys. Rev. Lett.",
    volume = "129",
    number = "1",
    pages = "011802",
    year = "2022"
}

@article{Gomes:2014yua,
    author = "Gomes, R. A. and Gomes, A. L. G. and Peres, O. L. G.",
    title = "{Constraints on neutrino decay lifetime using long-baseline charged and neutral current data}",
    eprint = "1407.5640",
    archivePrefix = "arXiv",
    primaryClass = "hep-ph",
    doi = "10.1016/j.physletb.2014.12.014",
    journal = "Phys. Lett. B",
    volume = "740",
    pages = "345--352",
    year = "2015"
}

@article{Choubey:2017dyu,
    author = "Choubey, Sandhya and Goswami, Srubabati and Pramanik, Dipyaman",
    title = "{A study of invisible neutrino decay at DUNE and its effects on $\theta_{23}$ measurement}",
    eprint = "1705.05820",
    archivePrefix = "arXiv",
    primaryClass = "hep-ph",
    doi = "10.1007/JHEP02(2018)055",
    journal = "JHEP",
    volume = "02",
    pages = "055",
    year = "2018"
}

@article{Chakraborty:2020cfu,
    author = "Chakraborty, Kaustav and Dutta, Debajyoti and Goswami, Srubabati and Pramanik, Dipyaman",
    title = "{Addendum to: Invisible neutrino decay: first vs second oscillation maximum}",
    eprint = "2012.04958",
    archivePrefix = "arXiv",
    primaryClass = "hep-ph",
    doi = "10.1007/JHEP08(2021)136",
    journal = "JHEP",
    volume = "08",
    pages = "136",
    year = "2021"
}

@article{Choubey:2020dhw,
    author = "Choubey, Sandhya and Ghosh, Monojit and Kempe, Daniel and Ohlsson, Tommy",
    title = "{Exploring invisible neutrino decay at ESSnuSB}",
    eprint = "2010.16334",
    archivePrefix = "arXiv",
    primaryClass = "hep-ph",
    doi = "10.1007/JHEP05(2021)133",
    journal = "JHEP",
    volume = "05",
    pages = "133",
    year = "2021"
}

@article{Choubey:2017eyg,
    author = "Choubey, Sandhya and Goswami, Srubabati and Gupta, Chandan and Lakshmi, S. M. and Thakore, Tarak",
    title = "{Sensitivity to neutrino decay with atmospheric neutrinos at the INO-ICAL detector}",
    eprint = "1709.10376",
    archivePrefix = "arXiv",
    primaryClass = "hep-ph",
    doi = "10.1103/PhysRevD.97.033005",
    journal = "Phys. Rev. D",
    volume = "97",
    number = "3",
    pages = "033005",
    year = "2018"
}

@article{deSalas:2018kri,
    author = "de Salas, P. F. and Pastor, S. and Ternes, C. A. and Thakore, T. and T\'ortola, M.",
    title = "{Constraining the invisible neutrino decay with KM3NeT-ORCA}",
    eprint = "1810.10916",
    archivePrefix = "arXiv",
    primaryClass = "hep-ph",
    doi = "10.1016/j.physletb.2018.12.066",
    journal = "Phys. Lett. B",
    volume = "789",
    pages = "472--479",
    year = "2019"
}

@article{KM3NeT:2023ncz,
    author = "Aiello, S. and others",
    collaboration = "KM3NeT",
    title = "{Probing invisible neutrino decay with KM3NeT/ORCA}",
    eprint = "2302.02717",
    archivePrefix = "arXiv",
    primaryClass = "hep-ex",
    doi = "10.1007/JHEP04(2023)090",
    journal = "JHEP",
    volume = "04",
    pages = "090",
    year = "2023"
}

@article{Beccaria:2026ous,
    author = "Beccaria, Martina and Ternes, Christoph A.",
    title = "{Probing damping effects in neutrino oscillations with the first JUNO data}",
    eprint = "2606.13362",
    archivePrefix = "arXiv",
    primaryClass = "hep-ph",
    month = "6",
    year = "2026"
}

@article{SNO:2018pvg,
    author = "Aharmim, B. and others",
    collaboration = "SNO",
    title = "{Constraints on Neutrino Lifetime from the Sudbury Neutrino Observatory}",
    eprint = "1812.01088",
    archivePrefix = "arXiv",
    primaryClass = "hep-ex",
    doi = "10.1103/PhysRevD.99.032013",
    journal = "Phys. Rev. D",
    volume = "99",
    number = "3",
    pages = "032013",
    year = "2019"
}

@dataset{herrera_2023_10822316,
  author       = {Herrera, Yago and
                  Serenelli, Aldo},
  title        = {Standard Solar Models B23 / SF-III},
  month        = nov,
  year         = 2023,
  publisher    = {Zenodo},
  version      = {v1.2},
  doi          = {10.5281/zenodo.10822316},
  url          = {https://doi.org/10.5281/zenodo.10822316},
}

@article{Magg:2022rxb,
    author = "Magg, Ekaterina and others",
    title = "{Observational constraints on the origin of the elements - IV. Standard composition of the Sun}",
    eprint = "2203.02255",
    archivePrefix = "arXiv",
    primaryClass = "astro-ph.SR",
    doi = "10.1051/0004-6361/202142971",
    journal = "Astron. Astrophys.",
    volume = "661",
    pages = "A140",
    year = "2022"
}

@article{XLZD:2024nsu,
    author = "Aalbers, J. and others",
    collaboration = "XLZD",
    title = "{The XLZD Design Book: towards the next-generation liquid xenon observatory for dark matter and neutrino physics}",
    eprint = "2410.17137",
    archivePrefix = "arXiv",
    primaryClass = "hep-ex",
    doi = "10.1140/epjc/s10052-025-14810-w",
    journal = "Eur. Phys. J. C",
    volume = "85",
    number = "10",
    pages = "1192",
    year = "2025"
}

@article{PandaX:2026qdp,
    author = "Chen, Peiyuan and others",
    collaboration = "PandaX",
    title = "{Measurement of solar $pp$ neutrino flux with the new PandaX-4T data}",
    eprint = "2607.02405",
    archivePrefix = "arXiv",
    primaryClass = "hep-ex",
    month = "7",
    year = "2026"
}

@article{T2K:2023smv,
    author = "Abe, K. and others",
    collaboration = "T2K",
    title = "{Measurements of neutrino oscillation parameters from the T2K experiment using $3.6\times 10^{21}$ protons on target}",
    eprint = "2303.03222",
    archivePrefix = "arXiv",
    primaryClass = "hep-ex",
    doi = "10.1140/epjc/s10052-023-11819-x",
    journal = "Eur. Phys. J. C",
    volume = "83",
    number = "9",
    pages = "782",
    year = "2023"
}

@article{NOvA:2021nfi,
    author = "Acero, M. A. and others",
    collaboration = "NOvA",
    title = "{Improved measurement of neutrino oscillation parameters by the NOvA experiment}",
    eprint = "2108.08219",
    archivePrefix = "arXiv",
    primaryClass = "hep-ex",
    reportNumber = "FERMILAB-PUB-21-373-ND",
    doi = "10.1103/PhysRevD.106.032004",
    journal = "Phys. Rev. D",
    volume = "106",
    number = "3",
    pages = "032004",
    year = "2022"
}

@article{MINOS:2017cae,
    author = "Adamson, P. and others",
    collaboration = "MINOS+",
    title = "{Search for sterile neutrinos in MINOS and MINOS+ using a two-detector fit}",
    eprint = "1710.06488",
    archivePrefix = "arXiv",
    primaryClass = "hep-ex",
    reportNumber = "FERMILAB-PUB-17-430-ND",
    doi = "10.1103/PhysRevLett.122.091803",
    journal = "Phys. Rev. Lett.",
    volume = "122",
    number = "9",
    pages = "091803",
    year = "2019"
}

@article{K2K:2006yov,
    author = "Ahn, M. H. and others",
    collaboration = "K2K",
    title = "{Measurement of Neutrino Oscillation by the K2K Experiment}",
    eprint = "hep-ex/0606032",
    archivePrefix = "arXiv",
    doi = "10.1103/PhysRevD.74.072003",
    journal = "Phys. Rev. D",
    volume = "74",
    pages = "072003",
    year = "2006"
}

@article{Super-Kamiokande:2006jvq,
    author = "Hosaka, J. and others",
    collaboration = "Super-Kamiokande",
    title = "{Three flavor neutrino oscillation analysis of atmospheric neutrinos in Super-Kamiokande}",
    eprint = "hep-ex/0604011",
    archivePrefix = "arXiv",
    doi = "10.1103/PhysRevD.74.032002",
    journal = "Phys. Rev. D",
    volume = "74",
    pages = "032002",
    year = "2006"
}

@article{Mishra:2023jlq,
    author = "Mishra, Nityasa and Strigari, Louis E.",
    title = "{Solar neutrinos with CE{\ensuremath{\nu}}NS and flavor-dependent radiative corrections}",
    eprint = "2305.17827",
    archivePrefix = "arXiv",
    primaryClass = "hep-ph",
    reportNumber = "MI-HET-812",
    doi = "10.1103/PhysRevD.108.063023",
    journal = "Phys. Rev. D",
    volume = "108",
    number = "6",
    pages = "063023",
    year = "2023"
}

@article{Barenboim:2020vrr,
    author = "Barenboim, Gabriela and Chen, Joe Zhiyu and Hannestad, Steen and Oldengott, Isabel M. and Tram, Thomas and Wong, Yvonne Y. Y.",
    title = "{Invisible neutrino decay in precision cosmology}",
    eprint = "2011.01502",
    archivePrefix = "arXiv",
    primaryClass = "astro-ph.CO",
    doi = "10.1088/1475-7516/2021/03/087",
    journal = "JCAP",
    volume = "03",
    pages = "087",
    year = "2021"
}

@article{FrancoAbellan:2021hdb,
    author = "Franco Abell{\'a}n, Guillermo and Chacko, Zackaria and Dev, Abhish and Du, Peizhi and Poulin, Vivian and Tsai, Yuhsin",
    title = "{Improved cosmological constraints on the neutrino mass and lifetime}",
    eprint = "2112.13862",
    archivePrefix = "arXiv",
    primaryClass = "hep-ph",
    reportNumber = "FERMILAB-PUB-21-779-T",
    doi = "10.1007/JHEP08(2022)076",
    journal = "JHEP",
    volume = "08",
    pages = "076",
    year = "2022"
}

@article{Escudero:2019gfk,
    author = "Escudero, Miguel and Fairbairn, Malcolm",
    title = "{Cosmological Constraints on Invisible Neutrino Decays Revisited}",
    eprint = "1907.05425",
    archivePrefix = "arXiv",
    primaryClass = "hep-ph",
    reportNumber = "KCL-2019-57",
    doi = "10.1103/PhysRevD.100.103531",
    journal = "Phys. Rev. D",
    volume = "100",
    number = "10",
    pages = "103531",
    year = "2019"
}

@article{Picoreti:2015ika,
    author = "Picoreti, R. and Guzzo, M. M. and de Holanda, P. C. and Peres, O. L. G.",
    title = "{Neutrino Decay and Solar Neutrino Seasonal Effect}",
    eprint = "1506.08158",
    archivePrefix = "arXiv",
    primaryClass = "hep-ph",
    doi = "10.1016/j.physletb.2016.08.007",
    journal = "Phys. Lett. B",
    volume = "761",
    pages = "70--73",
    year = "2016"
}

@article{Beacom:2002cb,
    author = "Beacom, John F. and Bell, Nicole F.",
    title = "{Do Solar Neutrinos Decay?}",
    eprint = "hep-ph/0204111",
    archivePrefix = "arXiv",
    reportNumber = "FERMILAB-PUB-02-061-A",
    doi = "10.1103/PhysRevD.65.113009",
    journal = "Phys. Rev. D",
    volume = "65",
    pages = "113009",
    year = "2002"
}

@article{Huang:2018nxj,
    author = "Huang, Guo-Yuan and Zhou, Shun",
    title = "{Constraining Neutrino Lifetimes and Magnetic Moments via Solar Neutrinos in the Large Xenon Detectors}",
    eprint = "1810.03877",
    archivePrefix = "arXiv",
    primaryClass = "hep-ph",
    doi = "10.1088/1475-7516/2019/02/024",
    journal = "JCAP",
    volume = "02",
    pages = "024",
    year = "2019"
}

@article{Valera:2024buc,
    author = "Valera, Victor B. and Fiorillo, Damiano F. G. and Esteban, Ivan and Bustamante, Mauricio",
    title = "{New limits on neutrino decay from high-energy astrophysical neutrinos}",
    eprint = "2405.14826",
    archivePrefix = "arXiv",
    primaryClass = "astro-ph.HE",
    doi = "10.1103/PhysRevD.110.043004",
    journal = "Phys. Rev. D",
    volume = "110",
    number = "4",
    pages = "043004",
    year = "2024"
}

\end{document}